\documentclass{aa} 

\usepackage[final]{changes}

\usepackage{graphicx}
\usepackage{hyperref}
\hypersetup{
    colorlinks=true,
    citecolor=blue,
    linkcolor=blue,
    filecolor=magenta,      
    urlcolor=blue,
}

\usepackage{txfonts}
\DeclareUnicodeCharacter{2212}{-}
\usepackage{natbib}
\bibpunct{(}{)}{;}{a}{}{,}

\def\kms  {km\,s$^{-1}$}

\def\bco {\ifmmode{^{12}{\rm CO}(J=2\to1)}\else{$^{12}{\rm
CO}(J=2\to1)$}\fi}

\def\mh     {H$_{2}$}

\def\sfr {M$_{\odot}$\,yr$^{-1}$}
\def\Lsun{L$_{\odot}$}
\def\Msun{M$_{\odot}$}
\def\deg {$^{\circ}$}

\def\arcsec {$^{\prime\prime}$}
\def\ppcm {cm$^{-2}$}
\def\pppcm {cm$^{-3}$}

\def\methanimine {CH$_2$NH}
\def\methanol {CH$_3$OH}
\def\Cyclopropenylidene{c-C$_3$H$_2$}

\begin{document}

\title{The opaque heart of the galaxy IC 860: Analogous protostellar, kinematics, morphology, and chemistry}

   \author{M. D. Gorski\inst{1}
          \and
          S. Aalto\inst{1}
          \and
          S. K{\"o}nig\inst{1}
          \and
          C. Wethers\inst{1}
          \and
          C. Yang\inst{1}
          \and
          S. Muller\inst{1}
          \and
          S. Viti\inst{2,3}
          \and
          J. H. Black\inst{1}
          \and
          K. Onishi\inst{1}
          \and
          M. Sato\inst{1}
          }

   \institute{Department of Space, Earth and Environment, 
            Chalmers University of Technology, Onsala Space Observatory, 
            439 92 Onsala, Sweden \\
            \email{mark.gorski@chalmers.se}
        \and
            Leiden Observatory, Leiden University, 
            P.O. Box 9513, 
            NL-2300 RA Leiden, The Netherlands \\
        \and
            Department of Physics and Astronomy,
            UCL, Gower Street, London, UK
             }

   \date{recived \today }


\abstract{

Compact  Obscured Nuclei (CONs) account for a significant fraction of the population of luminous and ultraluminous infrared galaxies (LIRGs and ULIRGs).
These galaxy nuclei are compact, with radii of 10-100~pc, with large optical depths at submm and far-infrared wavelengths, and characterized by vibrationally excited HCN emission. 
It is not known what powers the large luminosities of the CON host galaxies because of the extreme optical depths towards their nuclei.
CONs represent an extreme phase of nuclear growth, hiding either a rapidly accreting supermassive black hole or an abnormal mode of star formation.
Regardless of their power source, the CONs allow us to investigate the processes of nuclear growth in galaxies.
Here we apply principal component analysis (PCA) tomography to high-resolution (0\farcs06) ALMA  observations at frequencies 245 to 265~GHz of the nearby CON (59~Mpc) IC~860.
PCA is a technique to unveil correlation in the data parameter space, and we apply it to explore the morphological and chemical properties of species in our dataset.
The leading principal components reveal morphological features in molecular emission that suggest a rotating, infalling disk or envelope, and an outflow analogous to those seen in Galactic protostars.
One particular molecule of astrochemical interest is methanimine (\methanimine), a precursor to glycine,  three transitions of which have been detected towards IC~860.
We estimate the average \methanimine\ column density towards the nucleus of IC~860 to be $\sim10^{17}$cm$^{-2}$, with an abundance exceeding $10^{-8}$ relative to molecular hydrogen, using the rotation diagram method and non-LTE radiative transfer models. 
This \methanimine\ abundance is consistent with those found in hot cores of molecular clouds in the Milky Way.
Our analysis suggests that CONs are an important stage of chemical evolution in galaxies, that are chemically and morphologically similar to Milky Way hot cores.
}

  
\maketitle
\section{Introduction}
A significant portion of luminous and ultra-luminous (LIRGs and ULIRGs) infrared galaxies are shown to host nuclei that are compact (r$< 100$~pc),  hot (T $>$ 100~K), and opaque (N(H$_2$) $>$ 10$^{24}$~cm$^{-2}$). 
These compact obscured nuclei (CONs) account for approximately 40\% of ULIRGs and 20\% of LIRGs, yet it remains unknown what powers the extreme infrared luminosities of these galaxies \citep{Falstad2021}. 
The opaque nucleus may contain a supermassive black hole~(SMBH) with a high rate of accretion, or an abnormal mode of star formation \citep{Aalto2015b}.
How  galaxies grow SMBHs, and their relationship with the host galaxy, is one of the most fundamental questions in galaxy evolution \citep[e.g.,][]{Sanders1996, Ferrarese2000, Fabian2012}.
If CONs are powered by an active galactic nucleus~(AGN), then they likely represent a phase of rapid accretion onto the SMBH embedded in obscuring material.
The opaque nucleus is possibly due to a merger or interaction event \citep{Ricci2017,Boettcher2020}.
Regardless, CONs allow us to investigate the relationships between the nuclear growth processes in galaxies, SMBHs, and global galaxy properties. 

Compact obscured nuclei are currently identified by vibrationally excited HCN lines (HCN-vib;$\sum_{\rm{HCN−vib}}>1$\Lsun~pc$^{-2}$ in the J$=3-2$ transition) from the $\rm{v}_2=1$ state \citep[e.g.,][]{Salter2008,Sakamoto2010,Aalto2015a,Martin2016,Falstad2021}, and are opaque at frequencies above 80~GHz \citep{Barcos-Munoz2015,Sakamoto2017,Barcos-Munoz2018,Aalto2019}.
X-rays and mid-infrared wavelengths are strongly attenuated in these approximately $\rm{A_v}>1000$ galactic nuclei \citep[e.g.,][]{Treister2010,Roche2015}.
However, there are some indications that CONs may be identified using mid-infrared polycyclic aromatic hydrocarbon features and continuum ratios \citep{Garcia-Berente2022}. 
Nevertheless, the HCN-vib lines are probably excited by a strong mid-infrared radiation field.
This radiation field is likely due to a greenhouse effect where photons are trapped by dust.
This effect raises the dust temperature to of order 100~K \citep{Gonzalez-Alfonso2019}.
Thus, the CONs, at least partially, self-heat by trapping emission, similar to Galactic hot cores \citep{Kaufman1998}.

The similarities to Galactic hot cores do not end with the detection of vibrationally excited species and the trapping of infrared photons. 
Compact obscured nuclei are chemically rich in molecular lines. 
\citet{Martin2018} show that the spectrum between 218 and 363~GHz consists of transitions from at least 49 different species and $\sim1300$ transitions toward the known CON Arp~220. 
By comparison, hot cores may contain 10000 transitions between 329 and 363~GHz \citep{Jorgensen2016}, and the  ALMA Comprehensive High-resolution Extragalactic Molecular Inventory (ALCHEMI) has confirmed 1790 transitions between 84 and 373~GHz toward NGC~253 \citep{Martin2021}. 
Additionally, CONs seem to have strong, sometimes self-absorbed, lines of many molecules, such as HCN, HCO$^+$, H$_2$S, methanol~(CH$_3$OH), and CS  \citep{Aalto2015b,Aalto2019}. 
Of particular note is the detection of complex organic molecules (COMs), such as \methanol\ and methanimine~(\methanimine).
They exist in a variety of astrophysical environments (see reviews \citealp{Herbst2009,Caselli2012,Jorgensen2020}), but the largest body of detections of COMs comes from observations of hot cores \citep[e.g.,][]{Law2021}.
Toward these optically thick environments ($A_v>> 1$ \citealp{Ceccarelli1996}), COMs predominantly trace the optically thick, hot ($\sim$100--300 K), dense ($>$ 10$^7$ cm$^{-3}$) gas, in compact cores ($\sim$0.01--0.05~pc).
If COMs trace similar gas in galaxies, they may provide powerful diagnostics of the physics and chemistry of the innermost gas of CONs. 
Simply put,  CONs and hot cores contain hot dense material, have high optical depths, and are line-rich sources with many transitions from COMs. 

Methanimine (\methanimine) is a COM of particular interest in the interstellar medium, because its presence at high abundances ($\sim 10^{-8}$ relative to H$_2$) suggests that it is an important intermediate species in the growth of chemical complexity, especially among nitrogen-bearing species.
Typically, COMs are defined as having six atoms, while \methanimine\ has five.
However, \methanimine\ is the simplest ``imine'' and  a direct precursor of glycine \citep{Theule2011,Danger2011}, thus earning special status as a COM.
\methanimine\ was first detected in the interstellar medium toward Sagittarius B2 (Sgr B2).
Sgr B2 is one of the most massive, molecular clouds in the Milky Way with three embedded hot spots of star formation.
\methanimine\  is also the simplest molecule to contain the carbon-nitrogen double bond \citep{Godfrey1973}.

The first  detection of \methanimine\ in another galaxy was  with the Arecibo radio telescope toward Arp~220 \citep{Salter2008}.
Since then, \methanimine\ has been detected toward a number of galaxies \citep[e.g.,][]{Aalto2015b,Muller2011} and has been shown to exhibit maser activity at low frequencies toward CONs \citep{Gorski2021}.
\citet{Gorski2021} detected methanimine toward 80\% of CONs, with the only non-detection being toward IRAS~22491$-$1808, likely due to this galaxy's greater distance.
The excitation of \methanimine\ lines in CONs suggests this molecule is an especially valuable tracer of the CON environment. 

We employ radiative transfer models, rotation diagram analysis, and principal component analysis (PCA) to study the excitation of \methanimine\ and the morphology of the CON IC~860.
IC~860 is classified as a LIRG with log($L_{\rm IR}$ [\Lsun])=11.17; a distance of 59~Mpc has been determined through the use of H$_0=75\,{\rm km\, s}^{-1}\,{\rm Mpc}^{-1}$ and the adoption of a flat cosmology $\Omega_M=0.3$ and $\Omega_\lambda=0.7$ \citep{Sanders2003}. 
The only known closer CONs are ESO~320–G030 and NGC~4418 \citep{Sanders2003, Falstad2021}.
It is not known if IC~860 hosts an AGN and/or a starburst, as the nucleus is largely obscured by dust\citep{Alonso-Herrero2006}.
\citet{Aalto2019} investigated the nucleus of IC~860 with high angular resolution observations ($<$0.1\arcsec) with the Atacama Large Millimeter Array and the Very Large Array.
They uncovered evidence for a compact core fed by inflowing gas and a dense compact outflow. 
While the exact orientation and morphology of the nucleus remain unclear, \citet{Aalto2019} argue for a north-south oriented inflow and a molecular disk with an east-west oriented outflow (see Sect. 4.3.3 and Fig. 9). 
The inclination of the nuclear disk is rather unconstrained to between a ``near-face-on'' (i$\approx30^\circ$) and ``near-edge-on'' (i$\gtrsim60^\circ$) orientation.
Utilizing PCA as a statistical tool to study the spectrum presented by \citet{Aalto2019}, in addition to rotation diagram analysis and non-local thermodynamic equilibrium (non-LTE) radiative transfer modeling of the \methanimine\ lines, we offer a new penetrating look into chemically distinct  structures and excitation of molecular gas in the CON IC~860. 
We provide evidence that CONs have enhanced production of \methanimine\ compared to Galactic hot cores, and that the disk and outflow are oriented in a near-edge-on orientation.

\section{Observations}

The properties of the ALMA observations (ALMA project number 2016.1.00800.S) are reported by \citet{Aalto2019}, and the relevant details are summarized here.
Observations took place on November 17th 2017 with 45 antennas, and the amount of precipitable water vapor was $\sim1$~mm.
The maximal and minimal  baselines were 13.9~km and 0.113~km. 
The minimum baseline length of 113~m corresponds to a maximum recoverable scale of 2\farcs7. 
The bandpass calibrator, flux calibrator, gain calibrator, and check source are respectively J1256-0547, J1229+0203, J1327+2210, J1314+2348.
The  phase  center  was  set  to $\alpha$=13:15:03.5088  and $\delta$= +24:37:07.788  (J2000) and  the  data  were calibrated in the Common Astronomy Software Applications package \citep{McMullin2007}.

We utilize all four 1.875~GHz wide spectral windows centered at 246.4~GHz, 248.2~GHz 262.5~GHz, and 264.1~GHz.
The  original synthesized beam is $\sim0\farcs 05\times0\farcs 02$ with Briggs weighting parameter robust = 0.5. 
The final image cubes have been smoothed to a common beam of $0\farcs066\times 0\farcs041$ with a position angle of $20^\circ$ and contains 0\farcs010 pixels.

Channels are 8~MHz ($\sim 20$~\kms) wide.
All velocities are reported in the kinematic local standard of rest (LSRK) frame unless otherwise stated. 
The resulting sensitivities are 0.29~mJy, 0.32~mJy, 0.47~mJy, and 0.45~mJy  for the respective 246.4~GHz, 248.2~GHz 262.5~GHz, and 264.1~GHz.
Continuum subtraction was performed by means of the {\tt STATCONT} software package \citep{SanchezMonge2018}.
The peak fluxes for the 246.4~GHz, 248.2~GHz 262.5~GHz, and 264.1~GHz windows are $18.5\pm0.4$~mJy, $20.1\pm0.8$~mJy, $22.3\pm0.5$~mJy, and $19.9\pm1.0$~mJy.
The continuum was statistically derived on a pixel-by-pixel basis using the sigma clipping method for each individual spectral window.

\section{Results}
\begin{figure*}[]
    \centering
    \includegraphics[width=0.99\textwidth]{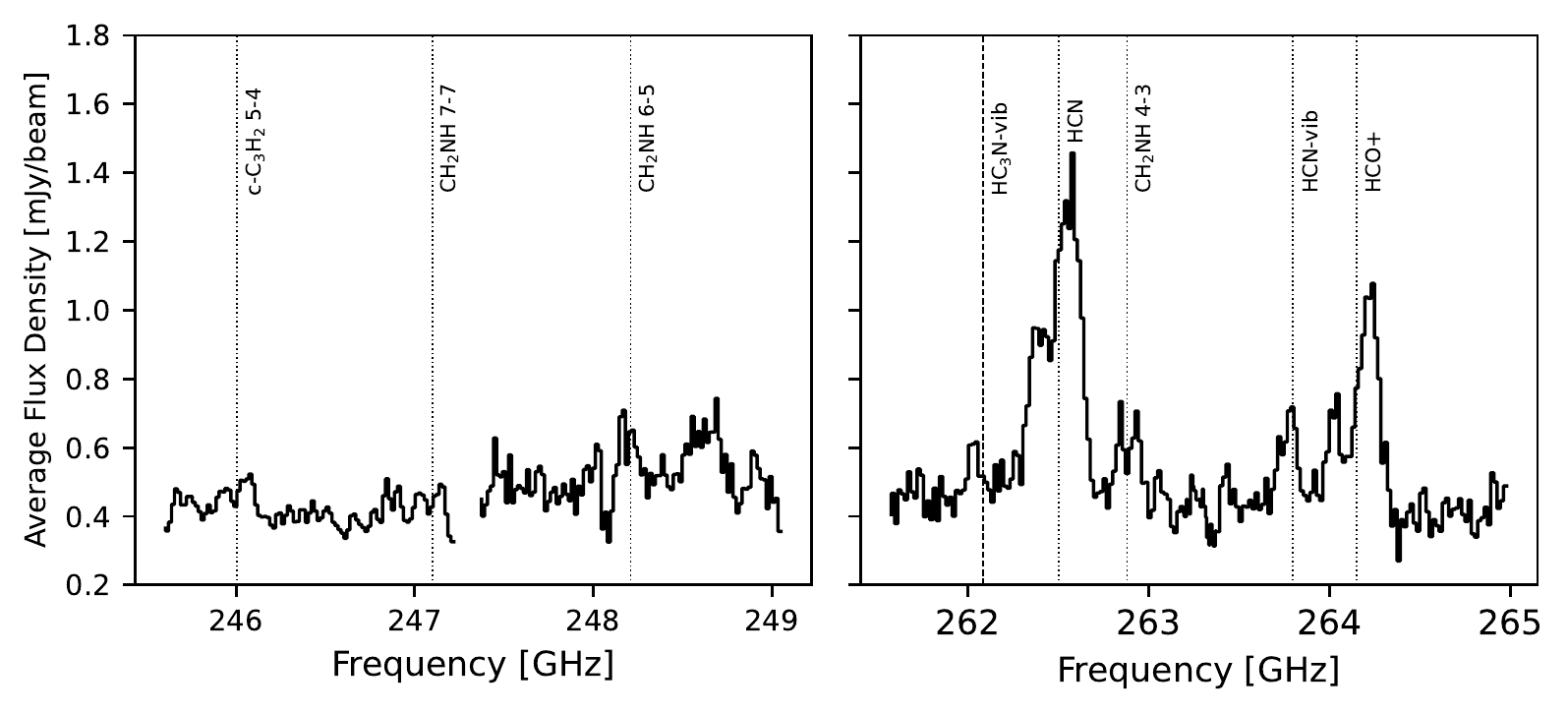}
    \caption{Average spectrum of the four image cubes created from the 1.875~GHz wide bands centered at 246.4~GHz, 248.2~GHz 248.2~GHz, and 262.5~GHz. Spectral lines of particular interest are labeled with vertical dotted lines.}
    \label{fig:avg}
\end{figure*}

\begin{figure*}[]
    \centering
    \includegraphics[width=0.99\textwidth]{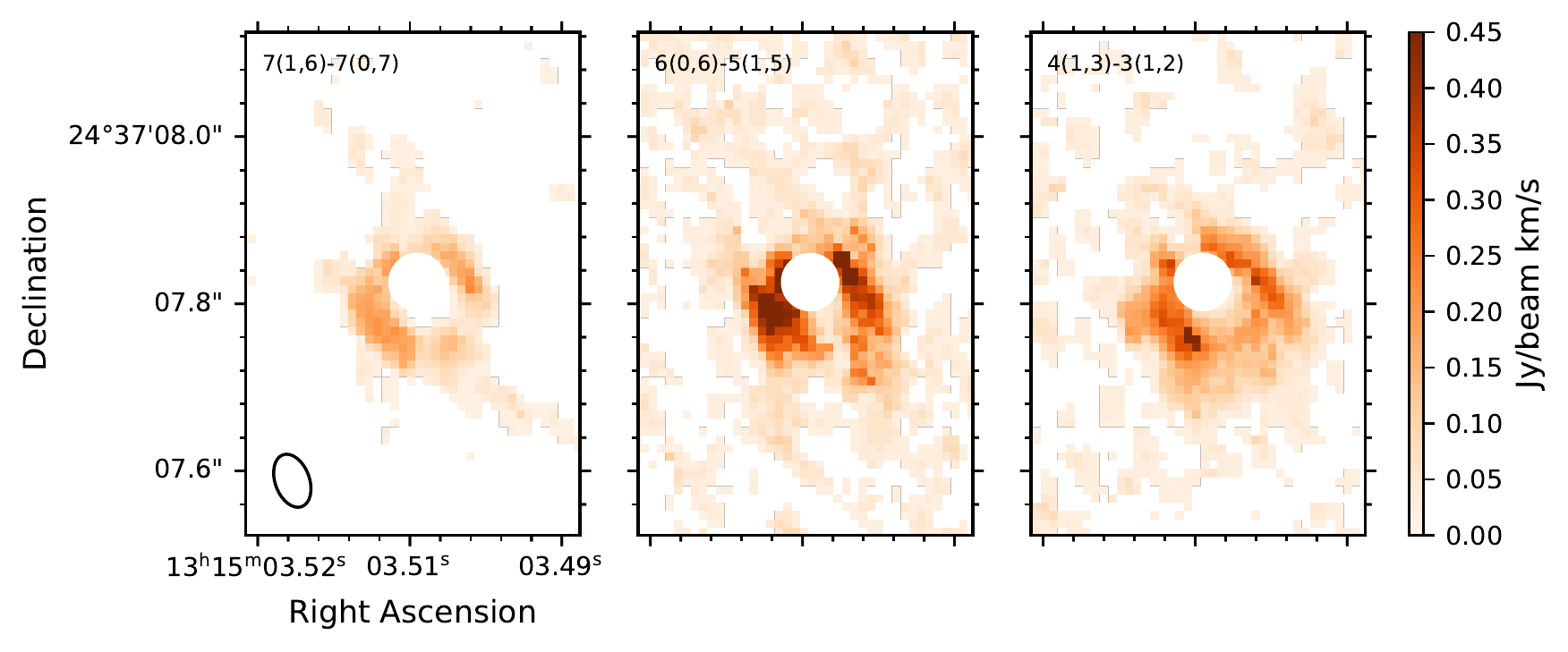}
    \includegraphics[width=0.99\textwidth]{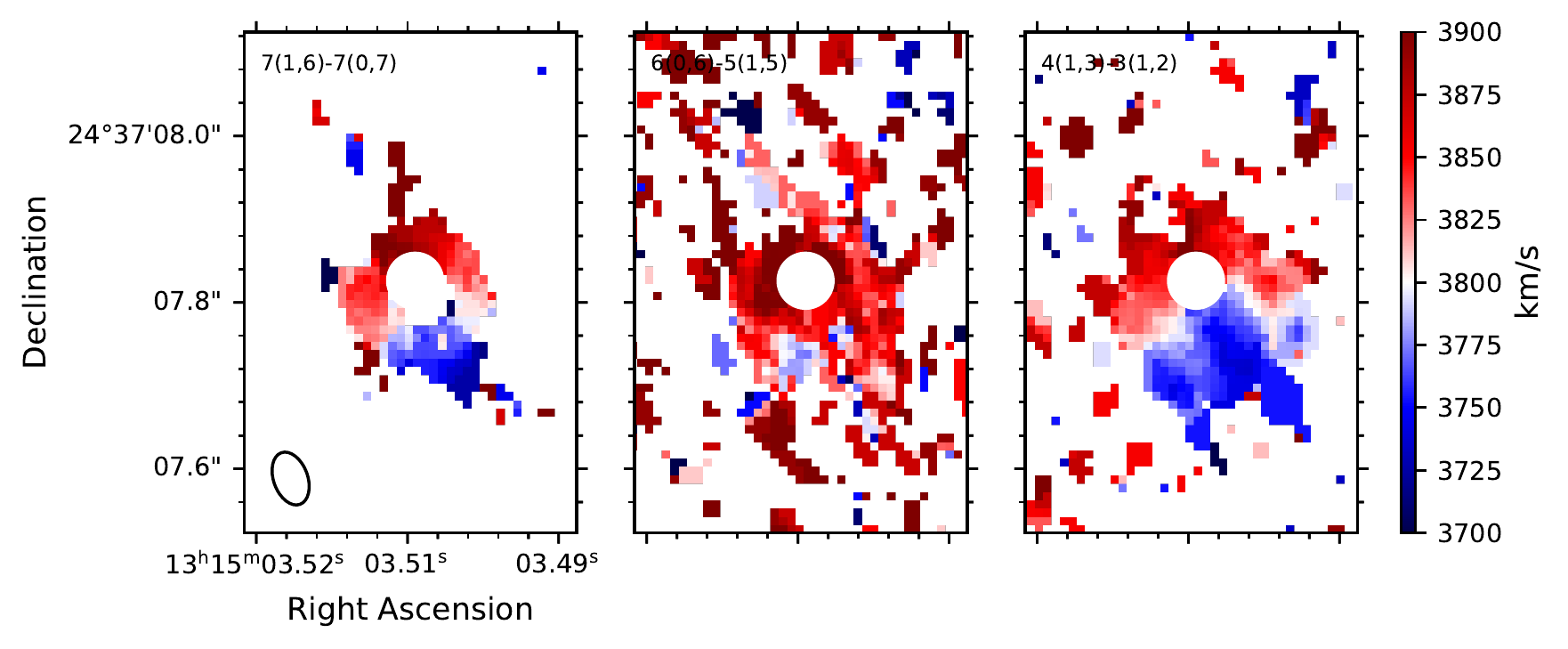}
    \includegraphics[width=0.99\textwidth]{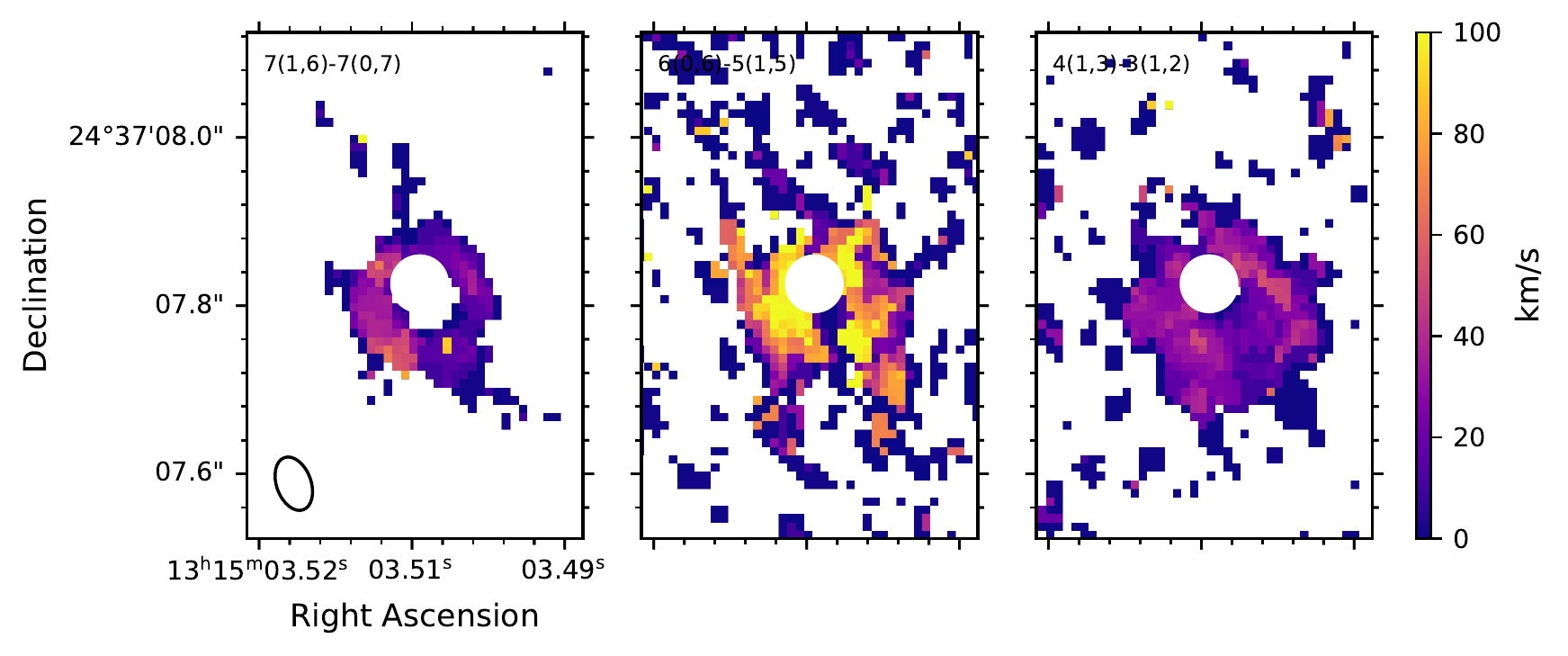}
    \caption{  Integrated intensity maps (top row), intensity-weighted velocity maps (middle row), and dispersion maps (bottom row) after removal of the continuum of the  \methanimine\  7(1,6)-7(0,7), 6(0,6)-5(1,5) and  4(1,3)-3(1,2) transitions. 
    Pixels with a signal $< \pm3$~times the sensitivity were excluded from the calculation of the velocity field, and the centermost 0\farcs07 is masked.
    }
    \label{fig:momen0}
\end{figure*}

Figure \ref{fig:avg} shows the average spectrum of the four image cubes made from the 1.875~GHz wide bands, and the spectroscopic data for relevant transitions are collated in Table~\ref{tab:spectro}.
Three transitions of \methanimine\ are detected in the nucleus of IC~860.
These are the 7(1,6)-7(0,7), 6(0,6)-5(1,5) and  4(1,3)-3(1,2) transitions, with respective rest frequencies at 250.1617~GHz, 251.4212~GHz, and 266.2700~GHz.
Figure \ref{fig:momen0} shows integrated intensity moment maps of the three identified transitions.
All  the \methanimine\ emission exists within a radius of 50~pc (0\farcs18) of the nucleus, and absorption is observed toward the innermost $\sim0\farcs1$.
The emission structure is similar to the HCN-vib emission reported in \citet{Aalto2019}, and is the strongest along the east-west axis perpendicular to the nuclear rotation. 

\begin{table}[]
    \centering
    \begin{tabular}{cccccc}
    \hline\hline
    Transitions  & Frequency & $E_u$ & $g_u$ & log($A_{ul}$) \\
                 & (GHz)     & (K)   & & (s$^{-1}$)        \\
    \hline
\Cyclopropenylidene\ $5_{23}$-$4_{32}$ & 249.0544 & 41.0 & 33 & $-3.38$\\
\Cyclopropenylidene\ $7_{06}$-$6_{16}$ & 251.3144 & 50.6 & 45 & $-3.03$\\
\Cyclopropenylidene\ $7_{06}$-$6_{16}$ & 251.3144 & 50.6 & 15 & $-3.03$\\
\Cyclopropenylidene\ $6_{15}$-$5_{24}$ & 251.5087 & 47.5 & 13 & $-3.17$\\
\Cyclopropenylidene\ $6_{25}$-$5_{14}$ & 251.5273 & 47.5 & 39 & $-3.17$\\
CH$_2$NH $7_{16}$-$7_{07}$ & 250.1618 & 97.2 & 15 & $-3.82$ \\
CH$_2$NH $6_{06}$-$5_{15}$ & 251.4213 & 64.1 & 13 & $-4.04$ \\
CH$_2$NH $4_{13}$-$3_{12}$ & 266.2700 & 39.8 &  9 & $-3.78$ \\
HCN $3$-$2$ & 265.8864 & 25.5 & 21 & $-3.08$ \\
HCN $\rm{v}_2 = 1$, $3$-$2$ & 267.1993 & 1050 & 21 & $-3.13$\\
HC$_3$N $\rm{v}_7 = 2$ J=29-28  & 265.4038 & 832.7 & 59 & $-2.83$ \\
HCO$^+$ $3$-$2$ & 267.5576 & 25.7 & 7 & $-2.38$\\ 
    \hline
    \end{tabular}
    \caption{Spectroscopic data for relevant transitions from Splatalogue (\url{https://splatalogue.online/}).}
    \label{tab:spectro}
\end{table}

All three transitions show a north-south velocity gradient. 
Figure \ref{fig:momen0} shows the velocity-weighed moment map including all pixels above $ 3\sigma$.
No negative fluxes are included, and therefore velocities toward the nucleus may not be reliable.
The center most 0\farcs07 is masked for this reason.
The dispersion maps (Fig.~\ref{fig:momen0}) generally agree with the HCN-vib maps shown by \citet{Aalto2019}.
Velocity dispersion ranges from $\sim20\rm{-}50$~\kms, with maxima along the east-west axis except for the 6(0,6)-5(1,5) transition which has a maximum dispersion of $>120$~\kms. 
The \methanimine\ 6(0,6)-5(1,5) line is contaminated toward the innermost 20~pc by cyclopropenylidene~(\Cyclopropenylidene) 
lines: an unresolved K-doublet of $J=7-6$ at 251.3143~GHz and a doublet of $J=6-5$ lines at 251.5087 and 251.5273~GHz, as seen in other CONs \citep{Sakamoto2021}. 
We extract the \methanimine\ spectra from the four pixels indicated in figure~\ref{fig:rotation_diagram}. 
Emission line parameters  from Gaussian fits (Table~\ref{tab:methanimine_lines}) show that along the east-west axis, the line widths of the 6-5 transition appear larger than the 4-3 and 7-7 transitions.
As a consequence, we do not use this line for abundance calculations toward these locations.

\begin{table}[]
    \centering
    \begin{tabular}{ccccc}
    \hline\hline
    Location & \methanimine\  & $T_{\rm{b}}$  & FWHM& $V_{lsr}$ \\
    & line & [K] & [\kms] & [\kms] \\
    \hline
      N & 4-3 & 13.4$\pm$3.2 & 53.7$\pm$15.0 & 3955$\pm$6\\
        & 6-5 & 22.0$\pm$5.0 & 33.6$\pm$8.5 & 3918$\pm$4 \\
        & 7-7 & 9.5$\pm$2.6 & 73.1$\pm$23.1 & 3938$\pm$10 \\ \hline
      S & 4-3 & 20.3$\pm$11.6 & 55.4$\pm$36.7 & 3795$\pm$16\\
        & 6-5 & -& - & - \\
        & 7-7 & 14.2$\pm$2.5 & 64.4$\pm$12.9 & 3813$\pm$5 \\ \hline
      E & 4-3 & 18.3$\pm$2.2 & 117.8$\pm$16.1 & 3862$\pm$7\\
        & 6-5$^b$ & 13.7$\pm$1.7 & 360.5$\pm$58.6 & 3921$\pm$23 \\
        & 7-7 & 13.4$\pm$1.7 & 119.9$\pm$17.1 & 3881$\pm$7 \\ \hline
      W & 4-3 & 19.7$\pm$2.4 & 97.4$\pm$13.6 & 3870$\pm$6\\
        & 6-5$^b$ & 15.5$\pm$2.4 & 141.6$\pm$25.4 & 3911$\pm$11 \\
        & 7-7 & 9.9$\pm$1.7 & 85.2$\pm$16.6 & 3865$\pm$7 \\ 
    \hline
    \end{tabular}
    \caption{\methanimine\ line parameters from Gaussian fits to the line profiles from the locations shown in Figure \ref{fig:rotation_diagram}.$^b$ The 6-5 transition is saddled by \Cyclopropenylidene transitions with rest frequencies of 251.5087, 251.5273, and 251.3143~GHz. The apparent line widths therefore appear larger toward the east and west locations, and these parameters are likely unreliable. Toward the north location, the 6-5 line width is consistent within the uncertainties of the 4-3 and 7-7 transitions.  }
    \label{tab:methanimine_lines}
\end{table}

\subsection{Methanimine abundances}

\begin{figure*}[]
    \centering
    \includegraphics[width=0.99\textwidth]{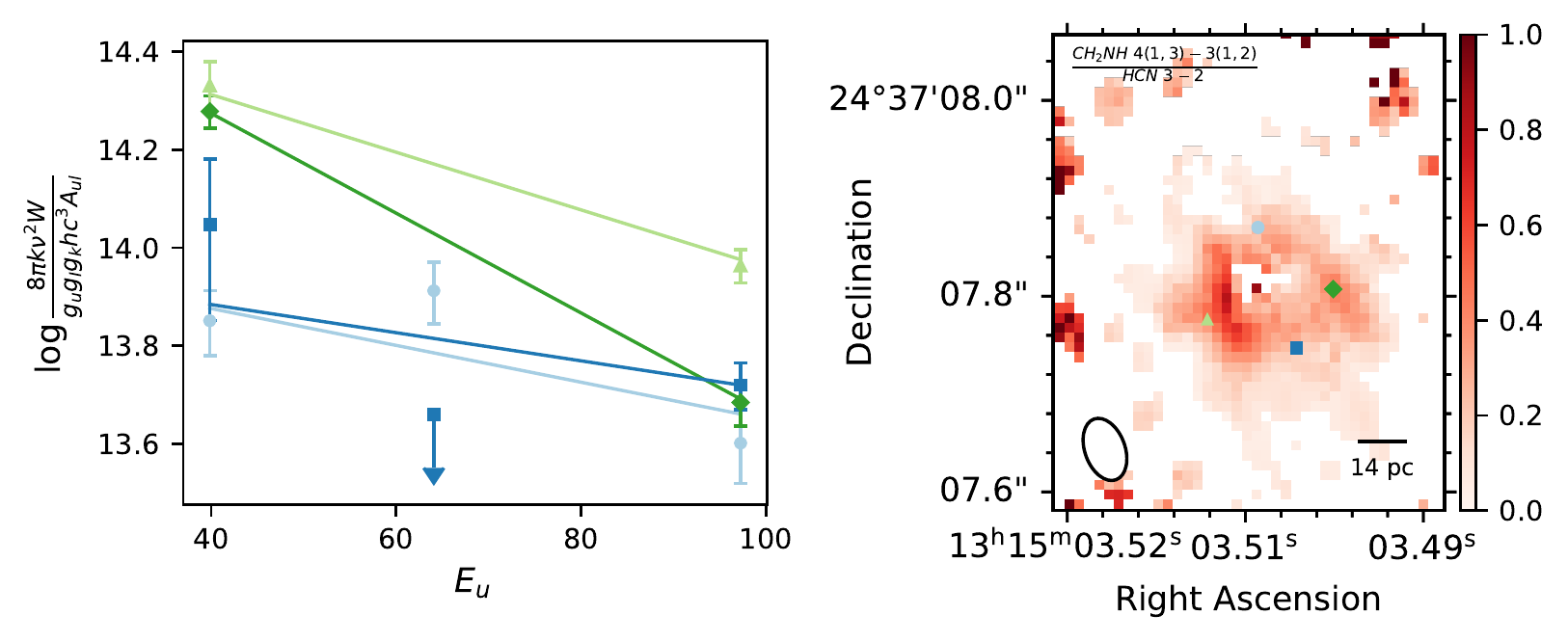}
    \caption{Rotation diagram for the \methanimine\ lines detected toward IC~860 (left) and the corresponding locations marked on the flux ratio map of the \methanimine~4(1,3)-3(1,2) to HCN~3-2 transitions (right). 
    The flux ratio map only includes pixels with a signal $> 3$~times the sensitivity of the \methanimine~4(1,3)-3(1,2) transition.
    }
    \label{fig:rotation_diagram}
\end{figure*}

We estimate column densities toward four locations in the innermost 100~pc of IC~860.
Four locations are chosen as points along the north-south and east-west axis (Fig.~\ref{fig:rotation_diagram}).
In either morphological scenario suggested by \citet{Aalto2019}, these four locations sample the outflow and disk. 


The rotation diagram method, or the population diagram method, is described in \citet{Turner1991} and \citet{Goldsmith1999}.
In short, one can estimate the column density and rotation temperature from a linear fit to the observed intensities and upper state energies, assuming local thermodynamic equilibrium.
\begin{equation}
\label{eq:turner1}
    \log \left ( \frac{3kW}{8\pi^3 \nu S \mu^2_i g_I g_K} \right) = \log \left (\frac{N}{Q_{ex}(T_{ex})} \right )-\frac{E_u}{k}\frac{\log e}{T_{ex}}
\end{equation}
Here the integrated intensity is $W$,  the intrinsic line strength is S,  the permanent electric dipole moment is $\mu$, $g_I$ and $g_K$ are the nuclear spin and K-level degeneracies, respectively,   the column density is $N$, $Q_{ex}$ is the rotational partition function, the upper-level energy is $E_u$, and $T_{ex}$ is the excitation temperature.
It is often easier to replace the inconvenient units of the $S \mu^2$, being Debye ($D$) or electrostatic units ($esu$), with the Einstein $A_{ul}$ coefficient in units of s$^{-1}$:
\begin{equation}
    A_{ul} = \frac{1}{g_u}\frac{64 \pi^4 \nu^3 S \mu^2}{3hc^2}.
\end{equation}
 Here $g_u=(2 J_u +1)$ is the rotation degeneracy of the upper level. 
 Then the left half of Eq.~\ref{eq:turner1} is much easier to work with in common units where W has units of K~\kms, $k$ is the Boltzmann constant in J~K$^{-1}$, $h$ is the Planck constant in J~Hz$^{-1}$, $N$ is in km$^{-2}$, and E$_u/k$  and $T_{ex}$ are in K:
 \begin{equation}
 \label{eq:turner2}
    \log \left( \frac{8 \pi k \nu^2 W} {g_u g_I g_k h c^3 A_{ul}} \right) = \log  \left( \frac{N}{Q_{ex}(T_{ex})} \right)-\frac{E_u}{k}\frac{\log e}{T_{ex}}.
\end{equation}
Here $g_I$ and $g_K$ are equal to one, as \methanimine\ is an asymmetric top and there are no interchangeable identical nuclei. 
The column density and rotation temperature are then the intercept and slope of a linear fit to the diagram.
We assume beam filling factors are unity and that the line emission is optically thin.
The ranges of excitation temperatures ($5$-$300$~K) and column densities($10^{15}$-$10^{19}$~cm$^{-2}$) were explored with the Markov chain Monte Carlo sampler {\tt emcee} \citep{Foreman2013}. 
Figure \ref{fig:rotation_diagram} shows the rotation diagram for four locations in IC~860, and the results of the fit are collated in Table~\ref{tab:rotation_diagram}.

\begin{table}[]
    \centering
    \begin{tabular}{cccc}
    \hline\hline
    Location  & $T_{\rm{b}}~[K]$  & $T_{ex}$~[K]& $\log($N(\methanimine) [cm$^{-2}])$ \\
    \hline
      N & 30.3 & $131_{-42}^{+80}$     & $17.22_{-0.14}^{+0.21}$ \\
      S & 22.7 & $151_{-72}^{+97}$     & $17.32_{-0.14}^{+0.21}$ \\
      E & 35.5 & $74_{-12}^{+20}$      & $17.43_{-0.04}^{+0.04}$ \\
      W & 30.4 & $43_{-4}^{+5}$        & $17.23_{-0.03}^{+0.03}$\\
    \hline
    \end{tabular}
    \caption{Rayleigh-Jeans brightness temperatures at 249~GHz, \methanimine\  excitation temperatures and column densities toward the locations indicated in Fig.~\ref{fig:rotation_diagram}. }
    \label{tab:rotation_diagram}
\end{table}

The average rotation temperature, measured by \methanimine\ emission, of IC~860 is $96_{-29}^{+44}$~K with an average column density of \methanimine\ is $2.0_{-0.3}^{+0.6} \times10^{17}$~cm$^{-2}$.
The emission line brightness temperatures are between a factor of two and an order of magnitude less than the derived excitation temperatures.
Consequently, the assumption of optically thin emission likely holds for the sampled regions.
\citet{Aalto2019} conclude that the continuum is a result of hot, 280 K dust. 
The continuum is likely more optically thin away from the opaque core toward the four regions identified in Fig.~\ref{fig:rotation_diagram}. 
We can estimate the optical depth from the observed intensity, assuming a beam filling factor of unity, from the relationship between optical depth and the temperature of a black body:
\begin{equation}
    \frac{I_\nu}{B_\nu(T_D)} = 1-e^{-\tau},
\end{equation}
where $I_\nu$ is the observed intensity, and $B_\nu(T_D)$ is the Planck function. 
For a brightness temperature of 30~K, $\tau$ is 0.1.
Following from \citet{keene1982} and \citet{Hildebrand1983}, the total gas column density is related to the optical depth by
\begin{equation}
    N({\rm H}+{\rm H_2})= 1.2\times10^{25}\,\tau\,(\lambda/400\mu {\rm m})^2  {\rm cm}^{-2},
\end{equation}
assuming H$_2$ dominates the gas and a standard dust-to-gas ratio of 1/100 \citep{Bohlin1978}, we find N($H_2$) $\approx 1.2\times10^{25}$~cm$^{-2}$.
The average \methanimine\ abundance (X[\methanimine]) is  thus $1.6\times10^{-8}$.

However, this method overestimates the masses toward opaque nuclei.
\citet{Wilson2014} and \citet{Aalto2019}  both show that these column density estimates toward Arp~220 and IC~860 respectively, overestimate the dynamical mass by factors of at least two.
Both groups effectively present a smaller dust-to-gas ratio as a means to rectify this discrepancy. 
For this reason, we present the derived \methanimine\ abundance as a lower limit.

\subsection{Principal component analysis tomography}

\begin{figure}
    \centering
    \includegraphics[width=0.5\textwidth]{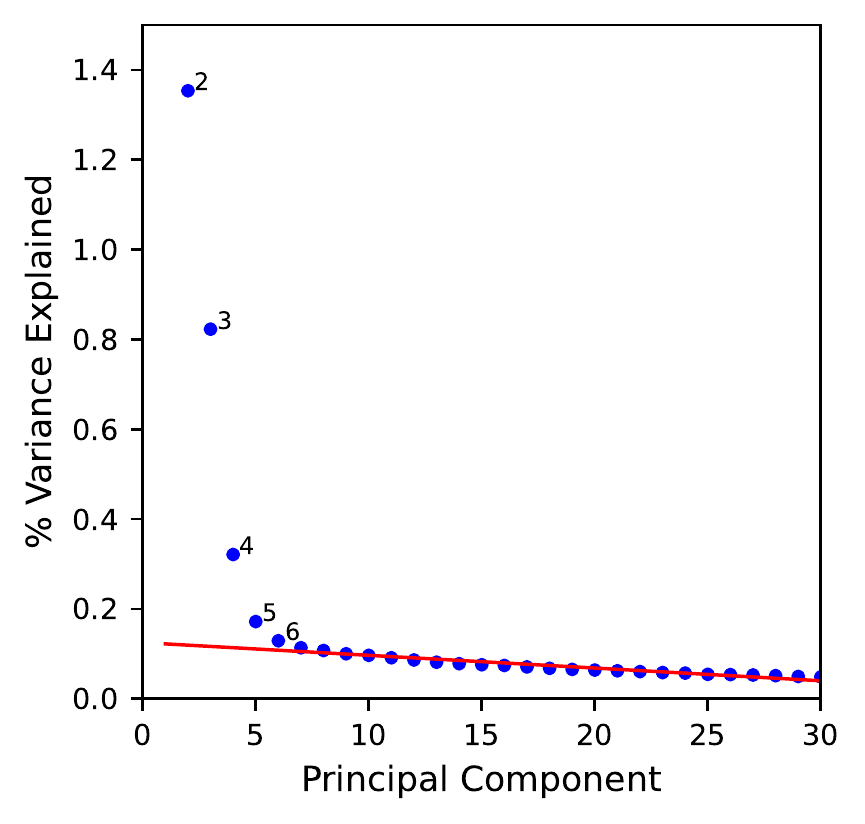}
    \caption{Scree test showing which PCs contain useful information.
    The red line shows a linear fit to PCs greater than 5.
    The first PC is out of scale, with a respective value of 0.923.
    From this test, there is useful information up to PC 6.
    }
    \label{fig:scree}
\end{figure}

In this section, we apply the technique of PCA to reveal morphological and chemical correlations in the data cubes.
PCA is the process of finding  eigenvectors of the data that describe orthogonal axes of decreasing variance, and is a form of dimensionality reduction.
Observations that contain many dimensions, for example two spatial, one frequency, and many molecules, benefit from dimensionality reduction.
The technique of PCA is commonly used to identify correlations between $n$ parameters in an  arbitrarily large sample of objects \citep[e.g.,][]{Murtagh1987,Heyer1997,Steiner2009}.
PCA is exceptionally useful for finding morphological correlations within spectral features in data cubes, and between images, in a variety of astronomical environments \citep[e.g.,][]{Meier2005,ScMuller2011,Ricci2011,Gratier2017,Navarete2021}.

\begin{figure*}
    \centering
    \includegraphics[width=0.995\textwidth]{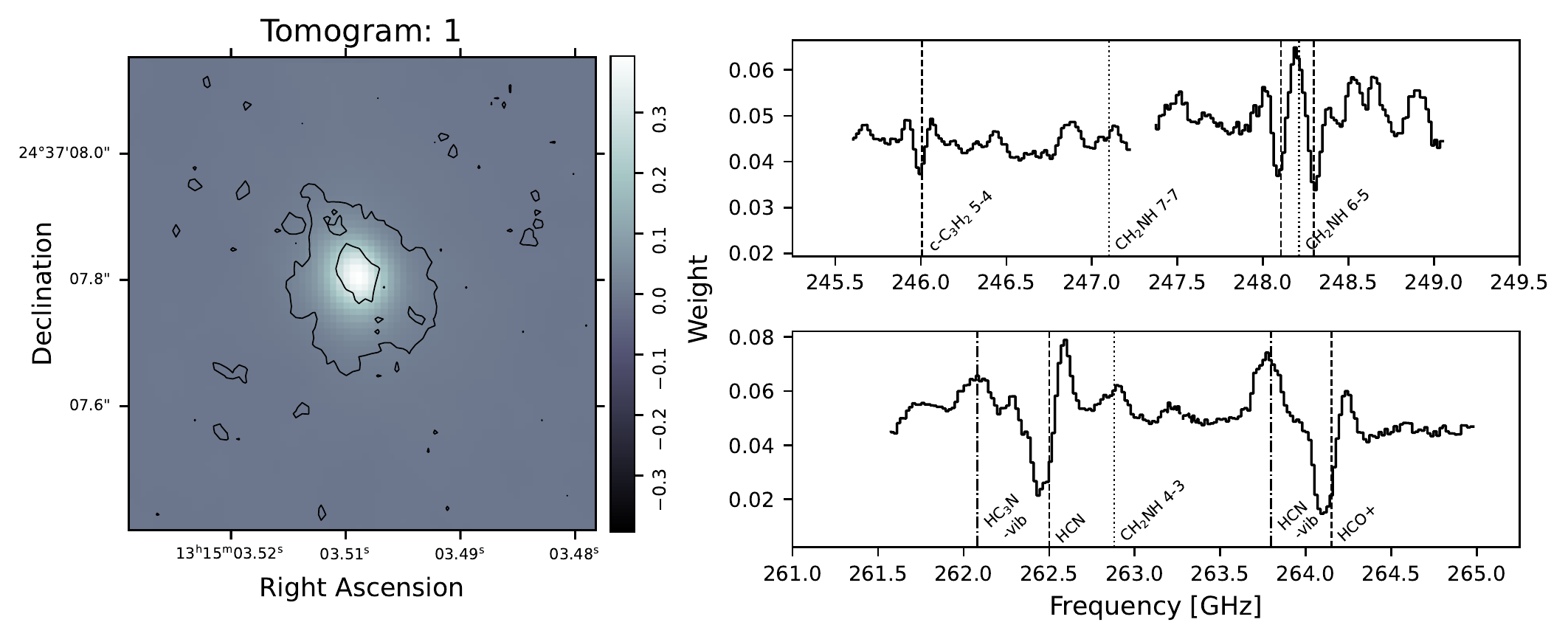}
    \caption{The first PC, which explains 92.29\% of the variance. The tomogram is plotted to the left and the eigenspectrum to the right. 
    White means stronger correlation, and black means anticorrelation.
    The 3$\sigma$ contour of the CH$_2$NH $4_{13}$-$3_{12}$ transition is plotted in black. 
    The vertical dotted lines represent the redshifted rest frequencies of the detected \methanimine\ transitions, the vertical dashed lines show the redshifted rest frequencies of the ground-state \Cyclopropenylidene, HCN, and HCO$^+$ lines, and the vertical dashed-dotted lines represent the redshifted  rest frequencies of the HCN-vib and HC$_3$N-vib lines.
    The two  lines on either side of the CH$_2$NH 6-5 transition are unlabeled and correspond to \Cyclopropenylidene\ transitions $\rm{J}=7-6$  and a doublet of $\rm{J}=6-5$ transitions.
    }
    \label{fig:PC0}
\end{figure*}

\begin{figure*}
    \centering
    \includegraphics[width=0.995\textwidth]{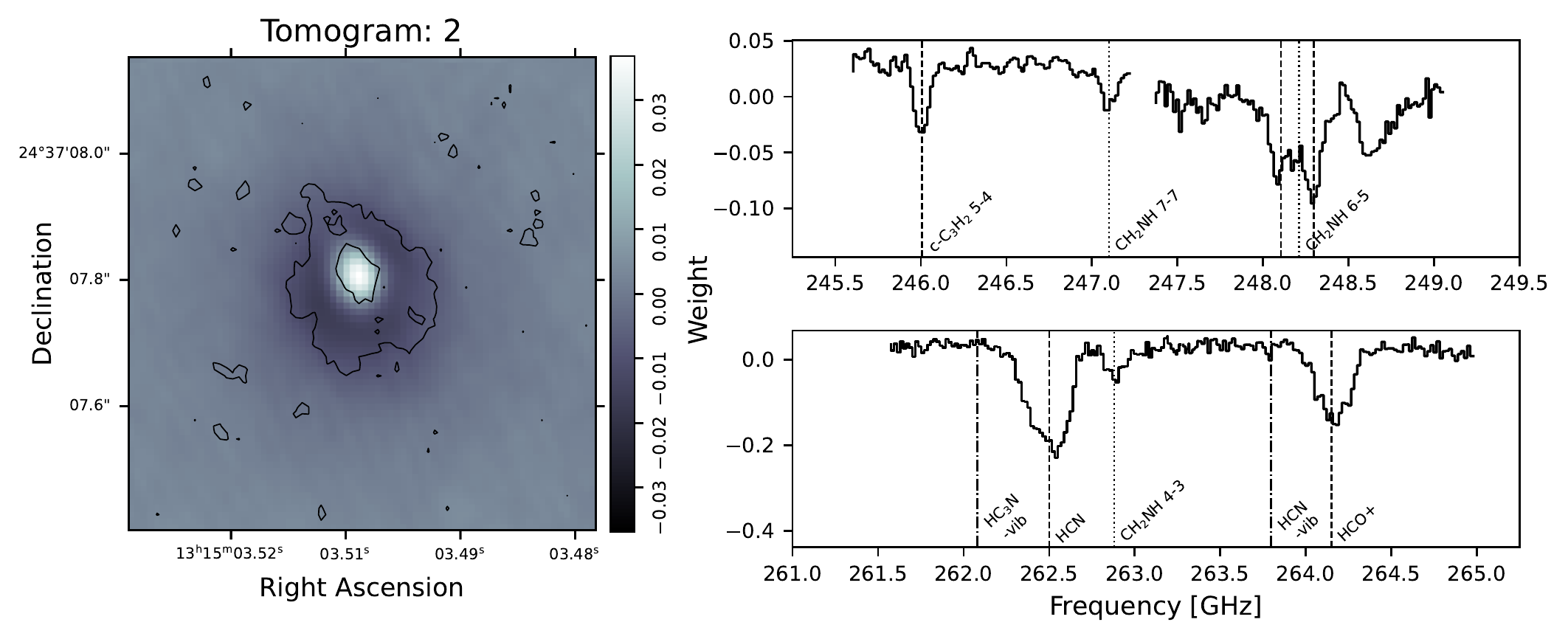}
    \includegraphics[width=0.995\textwidth]{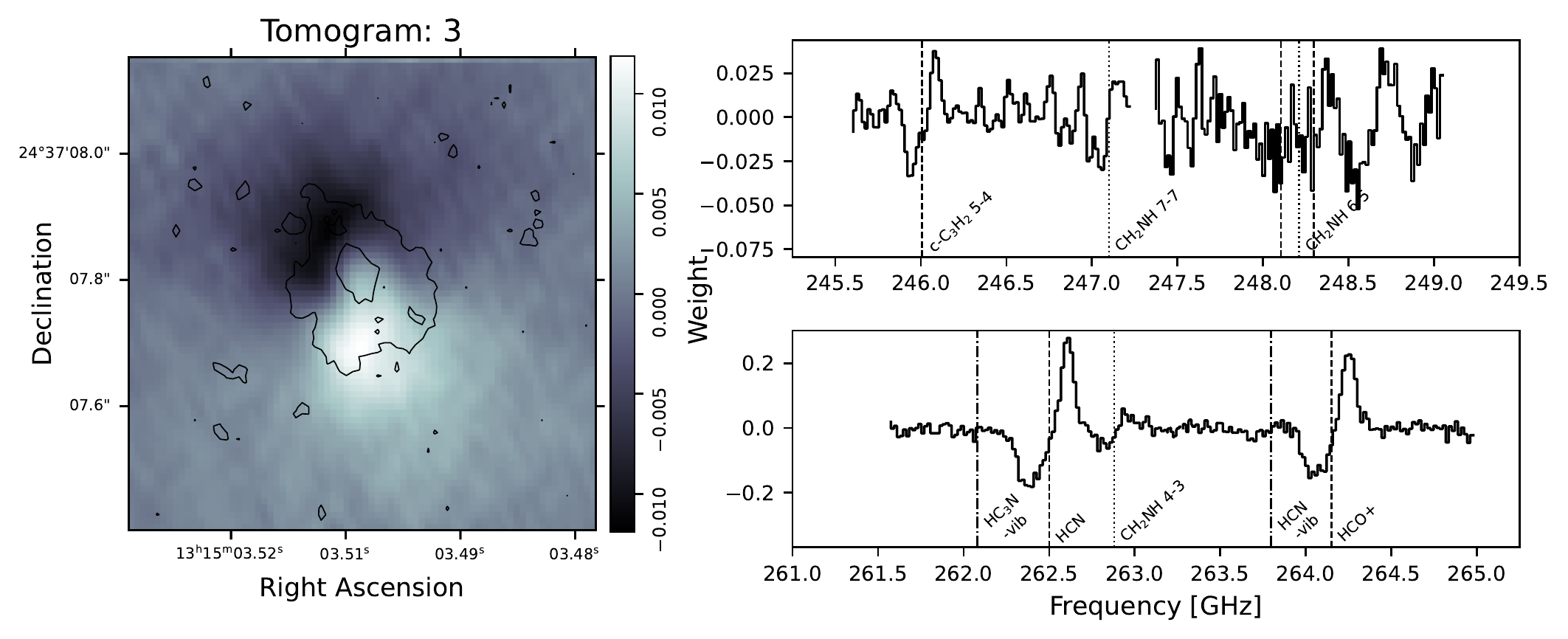}
    \caption{Principal components two and three. White means stronger correlation, and black means anticorrelation. 
    The 3$\sigma$ contour of the CH$_2$NH $4_{13}$-$3_{12}$ transition is plotted in black.
    PC 2 shows evidence for an ``envelope'' around the millimeter continuum, and PC~3 shows anticorrelated redshifted and blueshifted emission indicative of rotation.
    The vertical lines mark the redshifted rest frequencies of the same lines identified in Fig.~\ref{fig:PC0}.
    }
    \label{fig:PC1-3}
\end{figure*}

The method of PCA tomography implemented here is described by \citet{Steiner2009}.
PCA, when applied directly to the calibrated data cubes, returns the data in a system of uncorrelated orthogonal coordinates called principal components (PCs).
The first PC describes most of the variance within the data, with each subsequent PC describing less.
The total number of PCs is equal to the total number of variables.
Utilizing every PC it is possible to perfectly reconstruct the original data set.
For example, a data cube with 128 channels will perfectly be described by 128 PCs. 
However, the first few PCs usually contain the bulk of useful information, while the other PCs describe the less relevant information, such as uncorrelated information like noise. 
Each PC has a corresponding tomogram with an eigenspectrum.
The tomogram is a slice of the data in eigenvector space.
In short, PCA transforms the data into a new coordinate system:
\begin{equation}
    T_{\beta k}= I_{\beta \nu} \cdot E_{\nu k},
\end{equation}
where $\beta$ is the instructions to map the coordinates $x$ and $y$ to a two-dimensional matrix, $I_{\beta \nu}$ is the original data cube adjusted to have zero mean by subtracting the mean intensity for each frequency $\nu$, and $E_{\nu k}$ is the characteristic matrix.
The columns of  $E_{\nu k}$ correspond to eigenvectors also called eigenspectra.
$T_{\beta k}$ can be unmapped to $T_{x y k}$ to generate tomograms. 

We use the python toolkit Scikit-learn \citep{Pedregosa11a2011} to perform PCA on the IC~860 data cubes from \cite{Aalto2019}. 
The data cube is not normalized or scaled before PCA implementation.
The reason for this is we are interested in the relative strengths of the spectral lines.
Figure \ref{fig:scree} shows a scree test \citep{Cattell1966} for the PCs presented here, and we find that PCs below six contain relevant information, with noise dominating all other PCs.
 
The  first  PC (PC~1), shown in Fig.~\ref{fig:PC0}, explains  92.29\% of the variance. 
It shows correlations between the spectral features in the four data cubes and the continuum.
Reversed P-Cygni profiles are seen in HCN and HCO$^+$, and they are interpreted as evidence for self-absorption and inflowing molecular gas as in \citet{Aalto2015a}. 
Additionally, HCN-vib. HC$_3$N-vib, and \methanimine\ 4-3 and 7-7 lines appear to be correlated with the continuum, whereas the \methanimine\ 6-5 transition is confused with c-C$_3$H$_2$. 
This correlation with the HCN-vib and \methanimine\ lines is also observed in Zw049.057 \citep{Aalto2015b}.
The first PC has the benefit of identifying redundant information and is most similar to the integrated spectrum \citep[e.g.,][]{Ricci2011}.
This trend is seen across many galaxies \citep{ScMuller2011,Ricci2014}, where moving to higher-order PCs less variance is described, but nonredundant information is easier to identify.

PC~2 explains  1.35\% of the variance and is shown in Fig.~\ref{fig:PC1-3}.
Absorption lines of HCN, HCO$^+$, \methanimine, and \Cyclopropenylidene\ are strongly correlated with the millimeter continuum.
In an ``envelope'' around the strongest millimeter continuum, these lines are correlated with emission.

The PC~3 explains 0.82\% of the variance and is also shown in Figure~\ref{fig:PC1-3}.
It identifies anti-correlated redshifted and blueshifted emission in a symmetric structure about the center in the HCN, HCN-vib, HCO$^+$, \methanimine, and \Cyclopropenylidene\ lines.
The orientation of the major axis is north-south, with blueshifted emission in the south, and redshifted emission in the north. 
We interpret this structure as a rotating disk of material, and is similar to \citet{Heyer1997} who performed PCA on a modeled data cube of a ridged rotating cloud.
A major difference between \citet{Heyer1997} and the PC~3 presented here, is the asymmetric appearance of the positively and negatively correlated lobes. 
In the PC~3 tomogram, the positive and negative lobes appear swirled, implying noncircular motions. 
The HCN-vib line is weakly associated with this structure compared to the other ground-state transitions.
This is likely due to the HCN-vib transition tracing the innermost part of the nuclear region, identified in PC~4 discussed in the next paragraph. 
Similar structures are shown in NGC~7097 \citep{Ricci2011}, M81 \citep{ScMuller2011}, and ten other AGN \citep{Ricci2014}.
The vibrationally excited line HC$_3$N-vib is not identified in this structure.
Consequently, this  species likely traces some other significant structure. 

The fourth, fifth, and sixth PCs respectively explain 0.32\%, 0.17\%, and 0.13\%, of the variance.
The PC~5 identifies emission from HCN, HCO$^+$, methanimine,  cyclopropenylidene, HCN-vib, and HC$_3$N-vib in a double-lobed structure oriented in an east-west orientation.
We interpret this component as a biconical outflow.
The position angle of the outflow is 100\deg\ measured from the angle between the strongest pixels in the two lobes from PC~5. 
The resemblance to the outflow cone seen in the PCA of NGC~7097  from \cite{Ricci2011} is remarkable. 
\citet{Ricci2011} argue that this structure is only seen in the PCA of very edge-on systems. 
The presence of the vibrationally excited lines in the eigenspectrum of PC~5 also indicates that the excitation of these highly excited transitions is related to the outflow. 

The eigenspectra of PC~5 and 6, show anticorrelation between the low-velocity HCN and HCO$^+$ lines and their blue and red wings, suggesting these components are tracing an outflow or outflows.
PC~4 is significant in that it also shows anticorrelation between the redshifted and blueshifted features from HCN-vib, HC$_3$N-vib, and \Cyclopropenylidene~5-4 transitions, implying a second rotating disk within innermost 0\farcs2, or a change in the excitation of the molecular gas compared of PC~3.
The eigenspectrum of PC~4 also shows correlation with ground-state line wings of HCN and HCO$^+$ linking this rotating feature to the outflow.
The eigenspectrum of PC~6 shows correlation with the low velocity features of the ground state HCN line and weak anticorrelation with the line wings.
The fact that PC~4 and PC~6 show roughly opposite spatial orientations and slightly Doppler shifted eigenspectral features could be an indication of a rotating outflow. 
PC~6 is the least significant PC with a noisy eigenspectrum, thus we claim this is a tentative indication of a rotating outflow.

\section{Discussion}

\subsection{Non-LTE modeling}

The conditions within Compton thick nuclei are extreme.
Temperatures and molecular gas densities are respectively higher than $\gtrsim100$~K and $\gtrsim10^5$ cm$^{-3}$, and the molecular gas is exposed to intense continuous radiation from centimeter through near-infrared wavelengths (e.g., \citealt{Sakamoto2013},  \citealt{Aalto2015a},  \citealt{Aalto2015b},\citealt{Aalto2019}, \citealt{Mangum2019}, \citealt{Falstad2021}, \citealt{Gorski2021}). 
The simplifying assumptions that make the rotation diagram method possible may not yield accurate results under these conditions. 
\citet{Gorski2021} showed that an internal radiation field is a crucial factor in reproducing the observed excitation of \methanimine\ in CONs. 
In particular, they explored the excitation of the \methanimine\ 1(1,0)-1(1,1) transition, revealing a population inversion and thus maser emission.
However, in such an intense radiation field, the strongest transitions will be at millimeter and submillimeter wavelengths.

Using  {\tt GROSBETA}, described by \citet{Tabone2021} for its application to super-thermal OH emission, and by \citet{Gorski2021} to derive conditions for \methanimine\ masers, we model the observed \methanimine\ transitions toward IC~860.
GROSBETA is a non-LTE radiative transfer code and the successor to {\tt RADEX} \citep{vanderTak2007}.
The grid logarithmically samples densities $n$(H$_2$) $10^4\rm{-}10^9$~cm$^{-3}$ in 31 steps, column densities N(\methanimine) $10^{15}\rm{-}10^{19}$~cm$^{-3}$ in 17 steps, and linearly samples kinetic temperature from $10\rm{-}330$~K in 33 steps with a linewidth of 100~\kms.
We utilized crude collision rates scaled by dimensionless radiative line strengths to obtain order-of-magnitude estimates of the density in \methanimine-containing molecular gas \citep{Gorski2021}. A scaling factor was chosen so that the rate for the $1_{01} \to 0_{00}$ transition agreed with that computed by \citet{Faure2018}.
We also incorporate the radiation field utilized and described by \citet{Gorski2021}.
The  radiation field is constructed from  adopted observed fluxes as collected from the NASA/IPAC Extragalactic Database (NED)\footnote{https://ned.ipac.caltech.edu/, NED is funded by the US National Aeronautics and Space Administration and operated by the California Institute of Technology.} assuming that fluxes below frequencies of $\sim3\times10^{12}$~Hz are contained within the innermost 100~pc of IC~860.
We do not know what fraction of the total observed infrared flux is contained within the nucleus at subarcsecond resolutions.
Frequencies between $\sim3\times10^{12}$ and $\sim6\times10^{13}$~Hz are therefore scaled by a factor $\varphi$, where $\varphi$ is the fraction of observed infrared power contained within the source. 
We consider two model radiation fields consisting of two values of $\varphi$, 1.0 and 0.1.
Further details about the radiation field are explained in the Sect.~4.2 and the appendix by \citet{Gorski2021}.

\begin{figure*}
    \centering
    \includegraphics[width=0.98\textwidth]{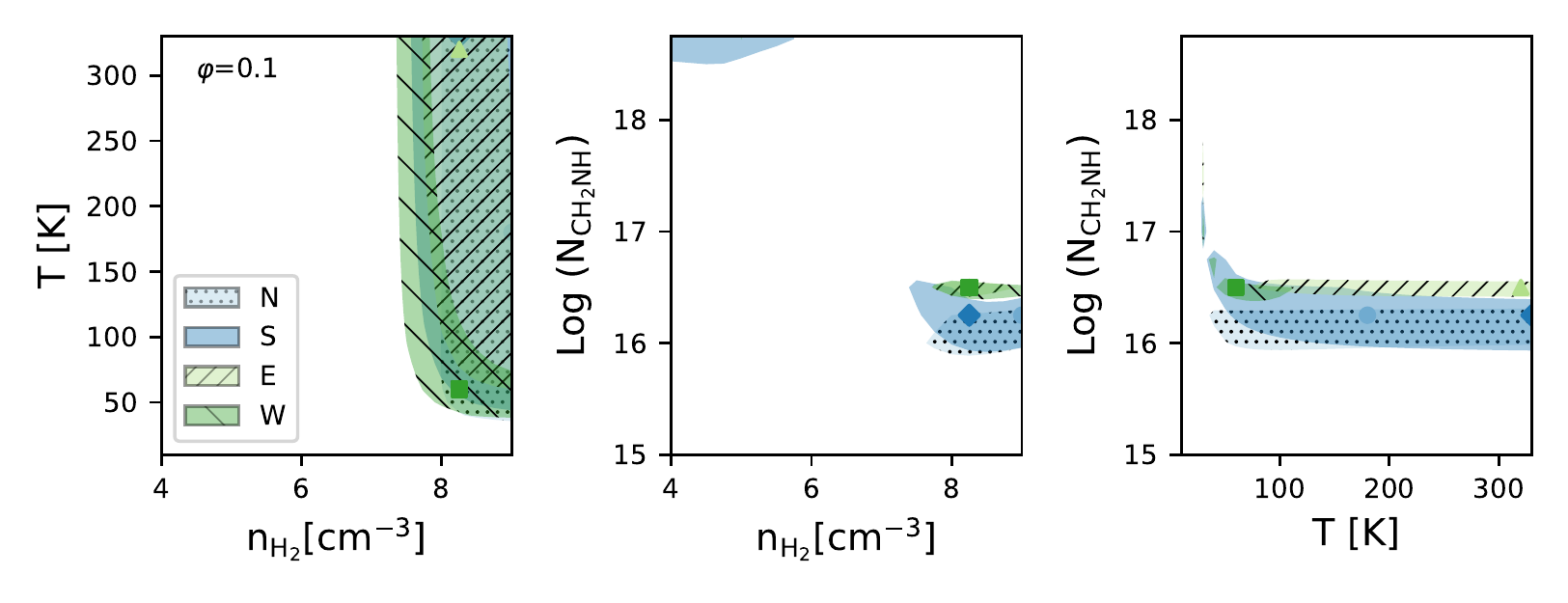}
    \includegraphics[width=0.98\textwidth]{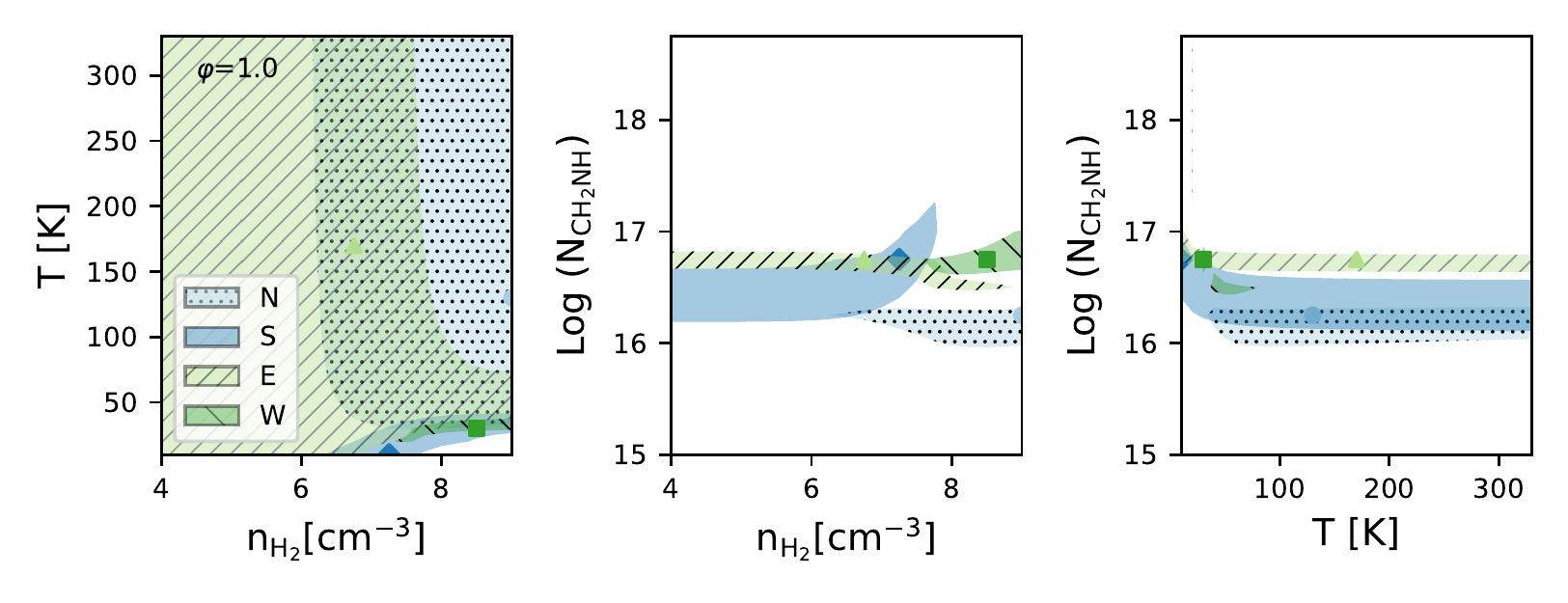}
    \caption{Non-LTE models showing the best fit to the detected \methanimine\ transitions. Blue contours represent the north-south axis, with the north region filled with dots. The east-west contours are shown in green with forward and backward hatch marks. The top row shows the $\varphi$=0.1 case and the bottom row shows the $\varphi=1.0$ case. The shaded area represents 3$\sigma$ from the best fit.In both cases, the temperature is generally unconstrained.
    In the $\phi=1.0$ case, the density is also unconstrained, but the  $\phi=0.1$ case has a lower limit of 10$^{7.5}$~\pppcm.
    Additionally, both cases show an enhancement of methanimine along the east-west axis. 
    }
    \label{fig:bestfit_contours}
\end{figure*}

The parameters of the best fits to the three observed \methanimine\ transitions are tabulated in Table \ref{tab:lvg_fits} and the 3$\sigma$ contours are shown in Fig.~\ref{fig:bestfit_contours}.
Blue contours represent the north-south axis, with the north region filled with dots. The east-west contours are shown in green with respective forward and backward hatch marks. 

The non-LTE models do not well constrain the temperature of the \methanimine-containing molecular gas, with uncertainties exceeding the almost entire range of the sampled parameter space.
Temperatures are unconstrained in the $\varphi=1.0$ case, and for $\varphi=0.1$ the kinetic temperature is $>30\pm10$~K.
Densities are also not well constrained in the $\varphi=1.0$ case, and for $\varphi=0.1$ case densities are generally $>10^7$~cm$^{-3}$. 
In both $\varphi$ cases, \methanimine\ column densities are also constrained by a lower limit of $10^{16.0}$~cm$^{-2}$. 

In general, the radiation field does not drastically change the estimate of the \methanimine\ column density.
Both $\varphi$ cases show a slight enhancement of \methanimine\ along the east-west axis.
However, in the $\varphi=0.1$ case, higher column densities up to $10^{19}$ are technically allowed within the $3\sigma$ contours. 
The results are in good agreement with the rotation diagram method with \methanimine\ column densities $\sim10^{17}$~cm$^{-2}$, as at current best the collision rates allow for order-of-magnitude precision \citep{Gorski2021}.
The average $\chi^2$ in the model fitting is lower in the $\varphi=1.0$ case, suggesting that the molecular gas toward these regions feels a significant fraction of the total observed infrared power of IC~860.
The models also corroborate more abundant methanimine along the east-west axis of IC~860 for all densities, temperatures, and radiation fields, though the precision of these models allows for significant overlap.

\citet{Aalto2019} claim that IC~860's nucleus consists of warm ($\sim280$~K) dense ($\sim10^7$~cm$^{3}$) molecular gas. 
Our models are consistent within the uncertainties with this interpretation.
The  non-LTE models  allow for \methanimine\ column densities two orders of magnitude greater than the rotation diagram method, illustrating that the radiation field may strongly influence the excitation of the molecular gas in IC~860.
Furthermore, both methods show evidence that \methanimine\ is more abundant along the east-west axis, suggesting that \methanimine\ is enhanced in the outflow of IC~860. 

\begin{table*}[]
    \centering
    \begin{tabular}{ccccc}
    \hline
    \multicolumn{5}{c}{$\varphi$ = 1.0} \\
    \hline
    Location & $\chi^2_{\rm{red}}$ & $T$~[K] & $\log(n(\rm{H}_2)$~[cm$^{-3}$]) & $\log($N(\methanimine) [cm$^{-2}])$ \\
    \hline
      N & 1.91 & 130 & 9.0 & 16.2\\
      S & 0.07 & 10 & 7.2 & 16.8\\ 
      \hline
      E & 0.01 & 170 & 6.8 & 16.8\\
      W & 0.60 & 30 & 8.5 & 16.8\\
    \hline
    \multicolumn{5}{c}{$\varphi$ = 0.1} \\
    \hline
      N & 2.44 & 180 & 9.0 & 16.2\\
      S & 0.03 & 330 & 8.2 & 16.2\\
      \hline
      E & 0.77 & 320 & 8.2 & 16.5\\
      W & 0.03 & 60 & 8.2 & 16.5\\
    \hline
    \end{tabular}
    \caption{Best fit \methanimine\ column densities, kinetic temperatures, and densities to the {\tt GROSBETA} grid toward the locations indicated in Fig.~\ref{fig:rotation_diagram}.$\varphi$ is the fraction of observed infrared power contained within the nucleus. The fit and uncertainties are plotted in Fig.~\ref{fig:bestfit_contours}. }
    \label{tab:lvg_fits}
\end{table*}

\subsection{Comparison to Milky Way environments}

\subsubsection{Abundances of methanimine in IC~860 and Milky Way environments}

Here we compare the \methanimine\ abundances in IC~860 to three environments known in the Milky Way:
The first is toward high-mass star-forming regions (HMSFRs), the second is the entire hot-core population of the Milky Way, and the third to Sgr B2. 
The average abundance of \methanimine\ in IC~860 relative to \mh, except toward the opaque core, to be approximately $10^{-8}$.
The abundance may be greater, as \mh\ column determinations are likely upper limits.

For Milky Way sources, \citet{Suzuki2016} report that the two most abundant \methanimine\ sources are HMSFRs Orion-KL and G10.47+0.03 with \methanimine\ abundances of 3.3$\times10^{-8}$ and 3.1$\times10^{-8}$ respectively. 
\citet{Neill2014} also report an abundance of  8.7$\times10^{-9}$ toward Sgr-B2 core. 
The abundance of \methanimine\ in IC~860 is similar to HMSFRs and Sgr~B2.
As \methanimine\ is not detected in low-mass star-forming regions \citep{Suzuki2016}, we assert that IC~860 is most similar to hot cores and HMSFRs.
The column densities and abundance of  \methanimine\ (this paper, \citealp{Suzuki2016}), the presence of HCN-vib emission \citep{Aalto2015a,Boonman2001}, and generally high molecular column densities ($\sim10^{25}~\rm{ cm}^{-2}$; \citealp{Aalto2019,Belloche2008}),  give credence to the idea that compact obscure nuclei may be scaled-up hot cores (i.e., on scales of tens of parsecs). 

\citet{Suzuki2016} argued that CH$_2$NH is formed mainly via radicals - NH + CH$_3$ -  that efficiently form in the gas phase once the mantles are evaporated.
These two radicals may be the product of the destruction of more complex species. We further speculate that indeed NH and CH$_3$ may be higher in strongly shocked gas (e.g. in the outflow region) because even a short period at very high temperatures  would lead to a fast destruction of NH$_3$ and CH$_4$. As shown in the chemical shock models in \citet{Viti2011}, for example, the reaction H + NH$_3$ is very efficient for shocks of velocities of at least 40 kms$^{-1}$, which can lead to a maximum temperature of $\sim$4000~K.
This formation story aligns well with the morphology of IC~860.
The inflow feeds the nucleus, where the molecular gas and dust is heated in the core of the CON. 
The warm and dense environment allows for gas phase reactions to happen, and the evidence is observed in the outflow.
We underline however that without a detailed chemical model of the gas observed in IC~860, which is beyond the scope of this paper, we can not confirm this picture. 

The high abundance of \methanimine\ in CONs could mean that as galaxies grow their nuclei, they also grow in chemical complexity.
We can compare a rough estimate of the mass of \methanimine\ contained in Milky Way hot cores and Sgr~B2 to IC~860.
\citet{Lintott2005} estimate that the total number of hot cores, molecular cores, and hot corinos in the Milky Way is $\sim10^4$. 
For the remainder of this estimate, these will be referred to as only hot cores. 
Generally, hot cores have radii between 0.01-0.05~pc \citep{Jorgensen2020, Law2021}.
If we assume that all hot cores have a large radius of 0.05~pc, spherical symmetry, and a column density of \methanimine\ of $(2.1\pm 0.7)\times10^{17}$~cm$^{-2}$ from the most \methanimine\ abundant hot cores \citep{Suzuki2016}, then the total amount of \methanimine\ in all the Milky Way in hot cores is $3.8\pm1.3$~\Msun.
Next, if we adopt  the average \methanimine\ column density  in IC~860 of $(2.0\pm 0.6)\times10^{17}$~cm$^{-2}$ from the rotation diagram  method (this paper),  the larger upper uncertainty as a symmetric uncertainty, and a radius of $\sim$60~pc \citep{Aalto2019}, the mass of \methanimine\ in the nucleus of IC~860 is $410\pm190$~\Msun.
This is  $110\pm60$ times worth all the \methanimine\ contained in the hot cores worth of the Milky Way. 
Adopting the  far UV (0.153~$\mu$m) + total infrared (8-1000~$\mu$m) star formation rate from \citet{Luo2022}  of $11.16\pm2.24$~\sfr, one would expect a mass of \methanimine\ of $43\pm17$~\Msun.
The mass of \methanimine\ is a factor of $12\pm6$ times greater in the nucleus of IC~860 compared to this expectation.
It is likely that this value is larger, as the calculation assumes large and \methanimine\ abundant hot cores. 

We also compare the column of \methanimine\ in Sgr~B2(N) to IC~860.
If we take the column density of \methanimine\ toward Sgr~B2(N), a hot spot of massive star formation, to be  $(7.0\pm 1.4)\times10^{16}$~cm$^{-2}$ \citep{Neill2014}, a radius of 2.3~pc \citep{Schmiedeke2016}, and circular symmetry, the  mass of \methanimine\ in Sgr~B2(N) is $0.27\pm0.03$~\Msun.
the core of IC~860 contains approximately $1900\pm500$ Sgr~B2(N)s worth of \methanimine. 
Considering, the complex organic nature of \methanimine,  CONs are potentially a significant manufacturer of complex chemistry in the Universe. 

From these rough estimates, one sees that IC~860 produces more \methanimine\ than predicted by star formation alone.
\citet{Aalto2019} suggest that IC~860 may be powered by a deeply embedded AGN or a super-Eddington starburst \citep[e.g.,][]{Andrews2011}.
IC~860's SMBH is predicted to be under massive, where the total enclosed mass of gas, dust, stars, and the SMBH is of order of the SMBH mass predicted by the $M_{\rm SMBH}-\sigma$ relationship ($M_{\rm SMBH}\sim10^7$ \Msun; \citealp{Aalto2019,McConnell2013,Davis2019}).
Less massive SMBHs also more commonly accrete at super-Eddington levels \citep{Shirakata2019,Farrah2022}.
Furthermore, IC~860's Pa$\alpha$ luminosity is under luminous for normal star formation \citep{Alonso-Herrero2006}. 
Hence, we find that it is probable that the power source of IC~860 is an embedded AGN.

\subsubsection{Structure and kinematics}

\begin{figure*}
    \centering
    \includegraphics[width=0.99\textwidth]{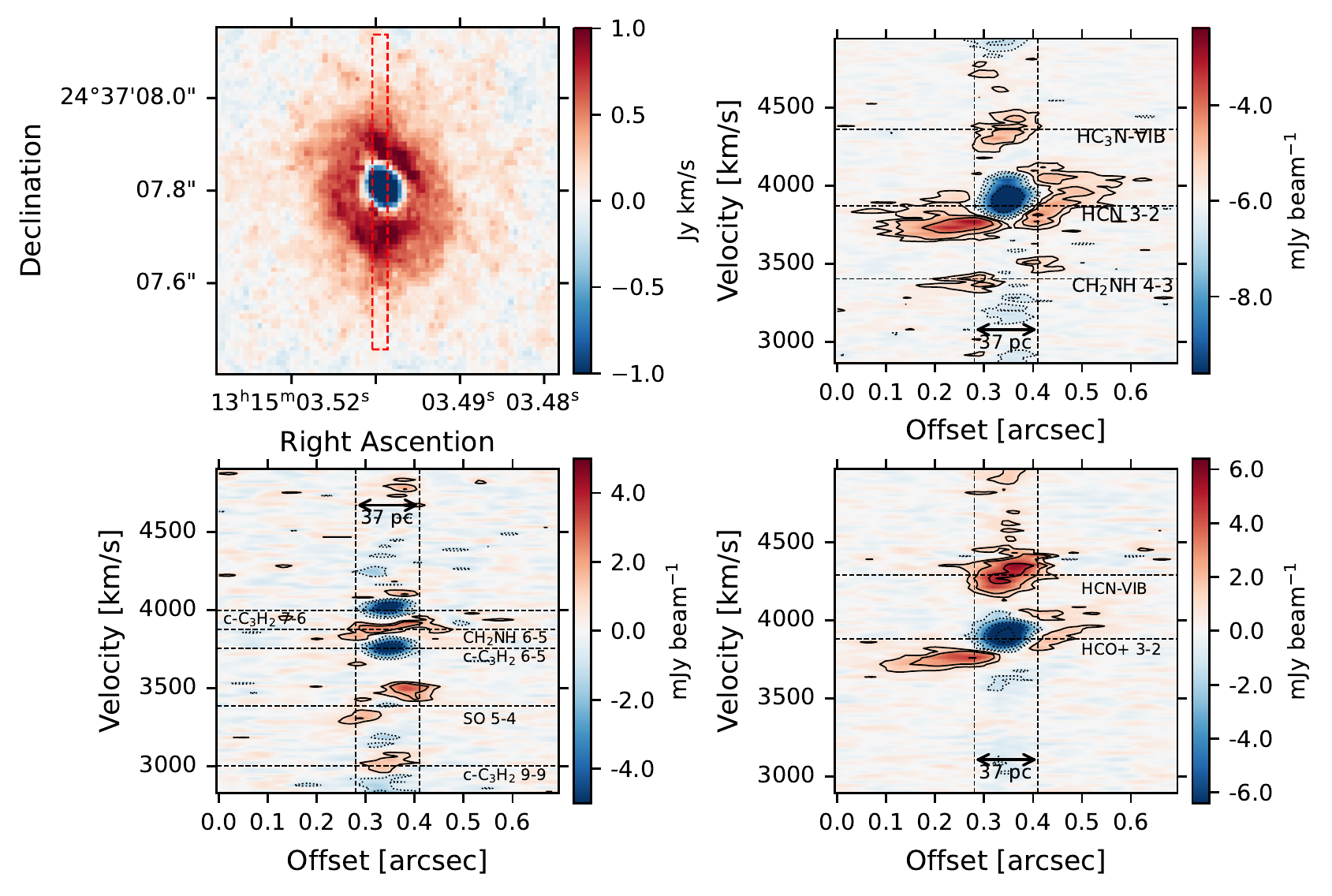}
    \caption{Position-velocity slice through the data cubes along the major axis of rotation.
    The top-left image shows the integrated flux map of the HCN 3-2 transition.
    The dashed red region shows the 0\farcs7 by 0\farcs03 slice through each data cube.
    The contours show 3, 6, 12, 24, and 48 times $\pm$0.3 mJy.
    We adopt line identifications from \cite{Sakamoto2021}, represented by horizontal black dashed lines; they show the systemic velocity of the indicated transitions.
    The velocity axis is shown in the optical convention utilizing the rest frequency of the HCN 3-2, \methanimine~6-5, and HCO$^+$~3-2 transitions.
    The  vertical  dashed black lines indicate the angular offset between the brightest redshifted and blueshifted velocity components.
    }
    \label{fig:PV}
\end{figure*}

\citet{Aalto2019} provide two potential scenarios for the nuclear structure of IC~860.
One scenario with a virtually face-on disk and a blueshifted outflow oriented almost directly toward the observer, and a second scenario with the outflow oriented perpendicular to the observer in the east-west orientation with an almost edge-on disk. 
With regard to the nuclear outflow, we argue in favor of the second scenario. 

From the PCA, we glean that the outflow is oriented in the east-west direction on the sky. 
The advantage of the PCA, over traditional methods of data cube analysis,  is that it can identify correlations over many spectral lines. 
PC~5 clearly shows a two-lobed structure oriented with a position angle of 100\deg\ on the sky, and the eigenspectrum shows correlated red and blue line wings in the HCN and HCO$^+$ transitions.
The vibrationally excited species HCN-vib and HC$_3$N-vib  show a second rotating structure in PC~4,
perhaps suggesting a correlation with the launching points of the outflow, a centrifugal barrier, or a nuclear disk surrounding the base of the outflow. 
From PC~3 alone, the envelope rotates in the north-south direction; however, there is significant width in the east-west direction, as indicated in the tomogram.
This could be due to a thick disk, flared disk, or more moderate inclinations. 

\citet{Aalto2019} also showed that the rotational major axis has a position angle of 0\deg\ and a peak rotational velocity of 100~\kms\ measured from the HCN-vib lines.
Position-velocity (PV) cuts along the major axis confirm that the rotational velocity is $\sim100$~\kms\ from the HCN-vib line (Fig.~\ref{fig:PV}). 
The HCN~3-2 emission is distributed over a much larger region of the sky, and the transition is seen in absorption toward the nuclear innermost $\sim$0\farcs1.

\begin{figure*}
    \centering%
    \includegraphics[width=0.99\textwidth]{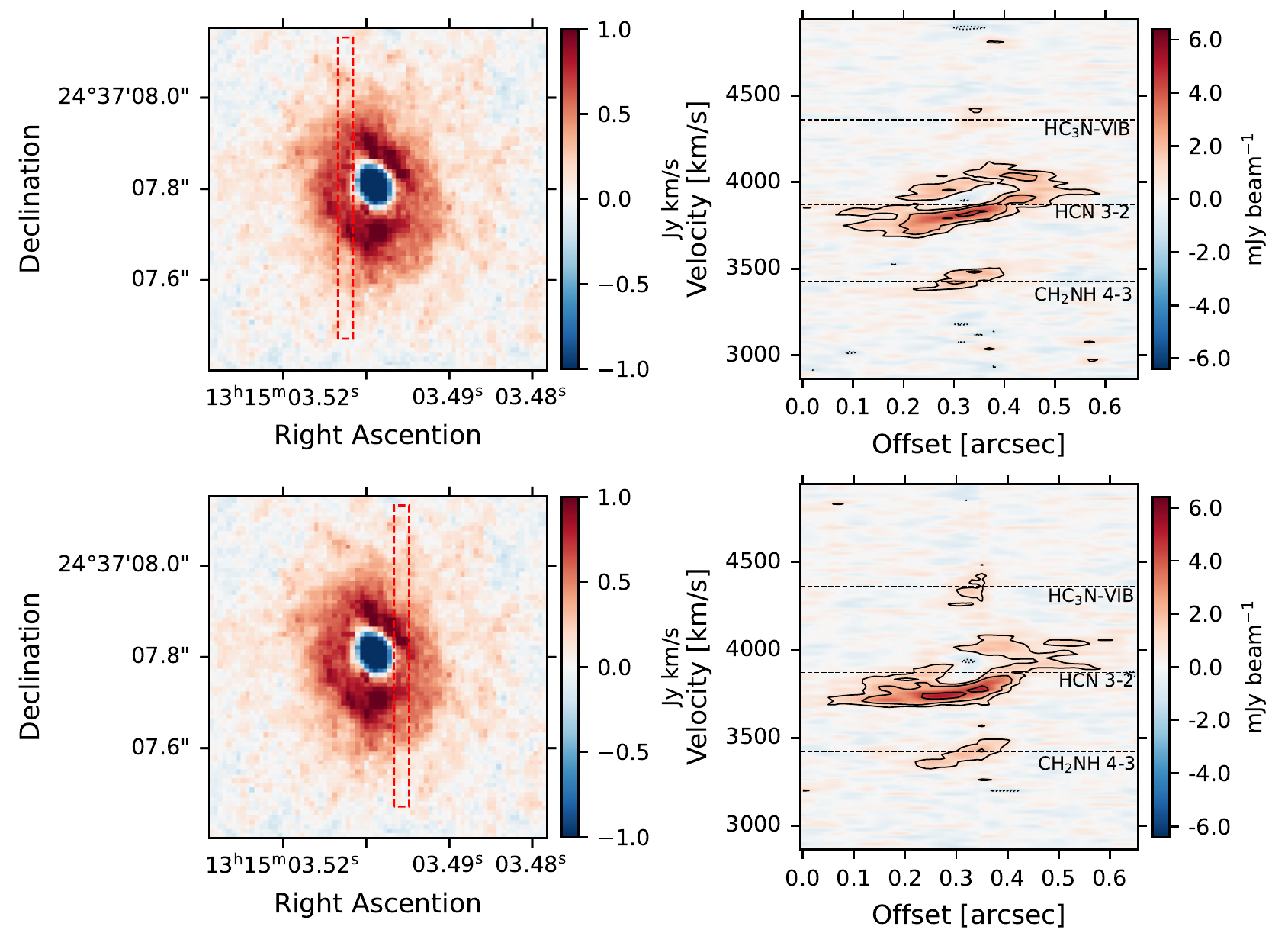}
    \caption{Position-velocity slices through the data cubes along the major axis of rotation. 
    The left column shows the integrated flux map of the HCN 3-2 transition, with the location of the 0\farcs7 by 0\farcs033  PV slice through the data cube containing the HCN 3-2 transition. 
    The contours show 3, 6, 12, 24, and 48 times 0.3 mJy.
    The  horizontal dashed black lines show the systemic velocity of the indicated transitions, and the velocity axis is shown in the optical convention utilizing the rest frequency of the HCN 3-2 transition.
    }
    \label{fig:PV-EW}
\end{figure*}

The PV diagrams show (Figure \ref{fig:PV}) that the brightest velocity components are redshifted and blueshifted, respectively, along the north-south axis, and the rotation signature disappears in the center.
The north axis also shows blueshifted emission, but there is a dearth of redshifted emission along the southern axis.
If the slice is shifted to the east, relative to rotation, there is clear redshifted emission (Fig.~\ref{fig:PV-EW}).
\citet{Sakai2014, Oya2014, Oya2016} argue these features are due to a rotating infalling envelope with a centrifugal barrier and an outflow, though on much smaller scales around protostars. 
Additionally, off the major axis PV plots that show such rotated elliptical structures are potential indicators of rotating outflows \citep{Tabone2020}. 
By a similar visual inspection, the angular offset between the redshifted and blueshifted velocity components, of the HCN~(3-2) line, is 0\farcs13 or $18$~pc.

\subsubsection{Feature enhancement and overall structure of IC~860}

Since PCA describes the data in eigenvectors, it is possible to re-weight and isolate these vectors such that they contribute more equally to the reconstructed image. 
It is important to remember that the eigenvectors are orthogonal and represent uncorrelated phenomena, and that noise can be reduced by eliminating eigenvectors that contain an excessive amount of noise. 
\citet{Steiner2009, Ricci2014} show that by normalizing each PC by its variance, removing noise dominated PCs, and reconstructing the data cube, one obtains an image cube where each orthogonal component contributes more equally to the final data product. 
This ``enhanced" image can reveal structures not easily discernible in the original data cube.
Since the CONs are optically thick in the center and contain many spectral lines, this method is attractive in the sense that it predicts structures that may not be directly observable. 
To clarify, we interpret the PCA-enhanced images as a reasonable extrapolation from the data, not an obvious detection of astrophysical phenomena.

\begin{figure*}
    \centering
    \includegraphics[width=0.98\textwidth]{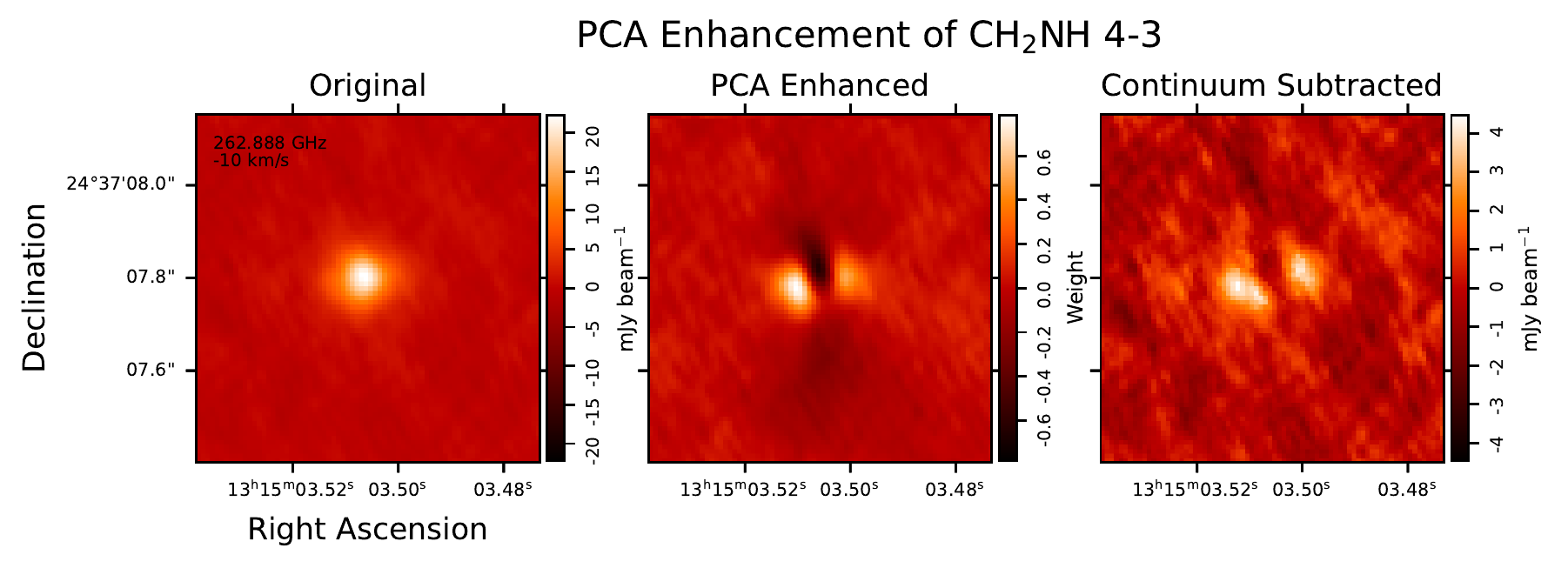}
    \caption{Comparison between the original data cube, the PCA-enhanced cube, and the continuum-subtracted data cube.
    Each panel shows the 262.888 GHz channel corresponding to -10~\kms (blueshifted) of the \methanimine~(4-3) transition, adopting a redshift of 0.0129 for IC~860. 
    The left panel displays the original data cube, the center the PCA-enhanced cube, and the right the continuum-subtracted cube.
    The center image is given in weight and the two other panels in mJy~beam$^{-1}$.
    }
    \label{fig:PCAE-methanimine}
\end{figure*}

Here we reconstruct the data cubes of IC~860 using PCs one through six. 
These components describe 94.9\% of the original data. 
Figure \ref{fig:PCAE-methanimine} shows the 262.888 GHz channel of each data cube centered on the \methanimine\ 4-3 line.
From left to right, the panels of Fig.~\ref{fig:PCAE-methanimine} show channels from the original data cube, the PCA-Enhanced data cube, and the continuum-subtracted data cube. 
Upon inspection, one sees that the original data cube is dominated by the millimeter continuum. 
The continuum-subtracted data cube shows a biconical structure oriented in the east-west direction and absorption toward the continuum. 
The PCA enhanced data cube reveals the same biconical structure, and absorption toward the nucleus, but these features are now equally weighted. 

Figures \ref{fig:PCAE-HCN} and \ref{fig:PCAE-HCO+} show selected blue and red channels of the HCN and HCO$^+$ transitions. 
The eigenspectra of PC~3 are largely dominated by these two lines, and thus the structure of the rotating disk is revealed more clearly. 
Each of these lines shows strong correlation with a narrow, less than one synthesized beam width (0\farcs06) wide, almost edge-on disk-like structure, and anticorrelation with the outflow.
We interpret this as evidence for a nearly edge-on nuclear disk.
If these predictions hold true, and the inclination of the nuclear disk is $\sim$90\deg, then the dynamical mass estimated by \citet{Aalto2019} is reduced to M$_{dyn}=2\times10^7$~\Msun.
We estimate the centrifugal barrier (Eqs. 3 and 4 \citealp{Sakai2014}) for a $2\times10^7$~\Msun\ object with a rotation velocity of 100$\pm20$~\kms\ is $17\pm7$~pc.
We adopt an ample 20~\kms\ uncertainty to the velocity, corresponding to the 8~MHz channel width.
This estimate is consistent with the 37~pc diameter offset seen in the PV diagrams, with an inclination $>61$\deg.

\begin{figure}
    \centering
    \includegraphics[width=0.5\textwidth]{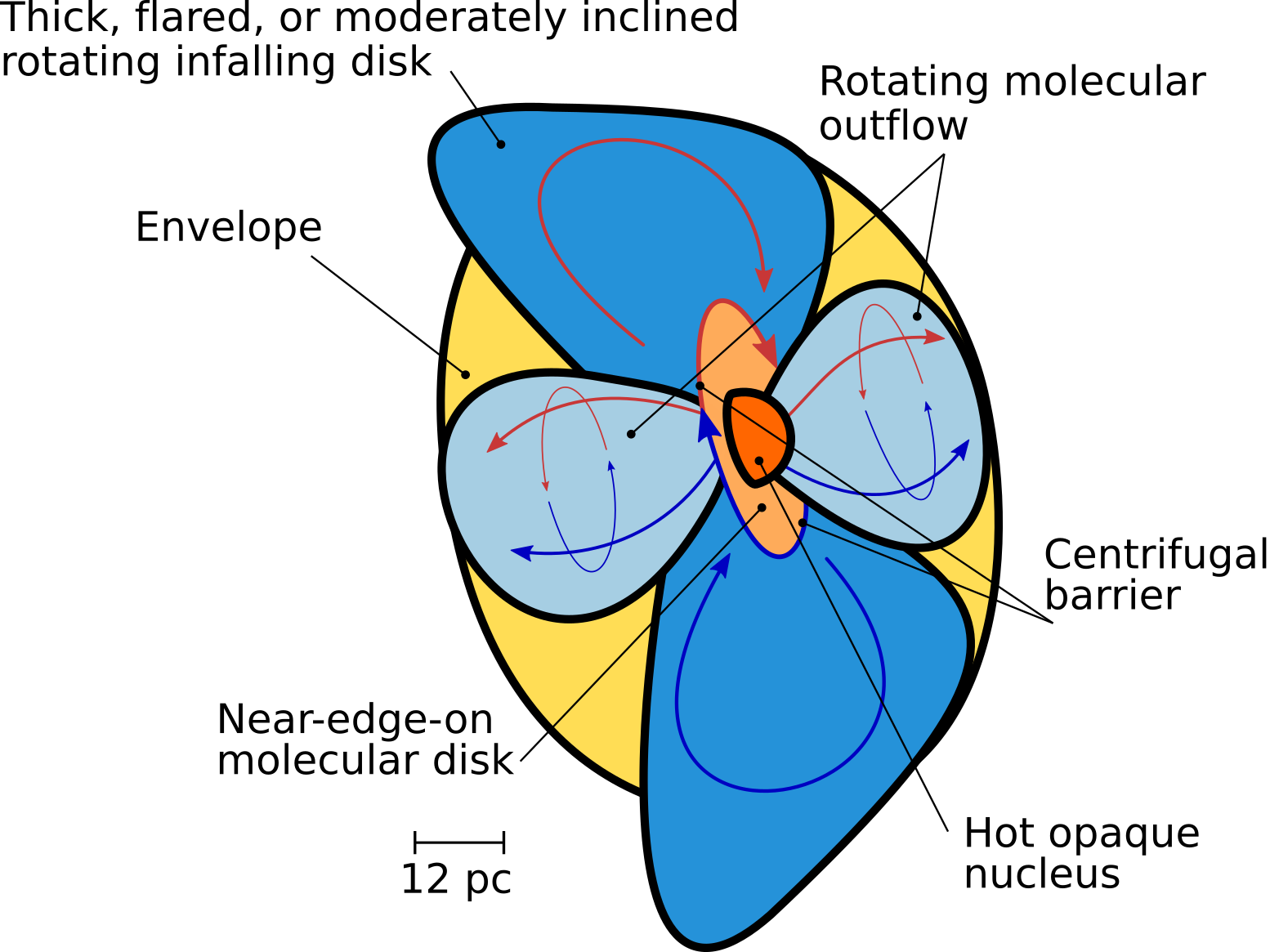}
    \caption{Schematic cartoon of the IC~860 ecosystem. 
    }
    \label{fig:cartoon}
\end{figure}

Examining all the PCs, the PCA-enhanced channel maps, and PV slices presented in this paper,  we construct a schematic cartoon of the IC~860 system (Fig.~\ref{fig:cartoon}). 
IC~860 consists of a flared, rotating, infalling molecular disk, and an encompassing envelope. 
The outflow is oriented east-west on the sky and shows evidence of rotation.
There is evidence of a highly inclined molecular disk.
In this picture, IC~860 has extraordinary similarities to the flared disk model of a rotating infalling envelope with a centrifugal barrier and an outflow around protostars \citep[e.g.,][]{Sakai2014}.

\section{Conclusions}

We have provided an analysis of ALMA 245-265~GHz 0\farcs06 resolution observations of molecular gas associated with the CON IC~860.
From PCA tomography, rotation diagram abundance determinations, and non-LTE radiative transfer models, we assert the following:

\begin{enumerate}

    \item Line ratio maps reveal that \methanimine\ is more abundant along the east-west axis of IC~860, and non-LTE models concur that the column density is enhanced along the outflow. 
    
    \item We calculated and modeled \methanimine\ column densities in IC~860  utilizing the rotation diagram method and non-LTE modeling. 
    Both methods yield \methanimine\ column densities ranging from $10^{17.2}$-$10^{17.4}$~\ppcm.
    The fractional abundances relative to \mh\ of \methanimine\ are $10^{-8}$ in good agreement with Milky Way HMSFRs.
    However, if we consider that the molecular (\mh) column density is probably overestimated, it is likely that methanimine is more abundant.
    Thus, IC~860 would have the highest fractional abundance of \methanimine\ currently known.
    
    \item The estimated total mass of \methanimine\ in IC~860 exceeds what would be predicted by star formation by a factor of $12\pm6$, hinting that CONs could be a significant producer of complex organic chemistry in the Universe. 
    
    \item Position-velocity slices through the data, and PCA, show that rotation slows at a radius of 18~pc, and there is a dearth of redshifted emission on the southern half of the major axis. 
     We interpret these features as the signature of a flared, rotating, infalling molecular disk analogous to those of Galactic protostars.
     
    \item We applied PCA tomography to our data cubes and explored the morphology of IC~860.
    Ground-state transitions of \methanimine, \Cyclopropenylidene, HCN, and HCO$^+$ trace a large rotating molecular envelope.
    Vibrationally excited species are revealed to be correlated with a second rotating structure in the innermost 0\farcs2.
    The outflow is revealed to be in the east-west orientation with an inclined molecular disk.
    Line wings of the ground-state lines of HCN and HCO$^+$ appear to be correlated with outflow, where \methanimine, appears to trace slower velocities.

\end{enumerate}
The nuclear power source of CONs is unknown due to their opaque nuclei. 
However, we have provided evidence that the abundance and mass of methanimine towards known CON IC~860 exceeds that predicted by star formation. 
Furthermore, there appears to be evidence for morphologically analogous structures similar to protostars.
Therefore, the growth process of CONs, and potentially SMBHs, could be analogous to that of Galactic hot cores or protostars, but acting at larger physical scales. 

\begin{acknowledgements}
M.G. would like to thank Professor Tiago Vecci Ricci for his helpful discussions on interpreting tomograms and eigenspectrums of data cubes. 

S.A., K.O., S.K., C.Y. gratefully acknowledge support from an ERC Advanced Grant 789410 a.
S.V acknowledge support from the European Research Council (ERC) under the European Union’s Horizon 2020 research and innovation program MOPPEX 833460.

M.G. acknowledges support from the Nordic ALMA Regional Center (ARC) node based at Onsala Space Observatory. 
The Nordic ARC node is funded through Swedish Research Council grant No 2017-00648

\end{acknowledgements}
\bibliographystyle{aa_url} 
\bibliography{methanimine,maser,Con Quest,misc,jb-add,pca,COMs} 


\begin{appendix}

\section{Principal components and eigenvectors, and PCA-enhanced data cubes}

\begin{figure*}
    \centering
    \includegraphics[width=0.99\textwidth]{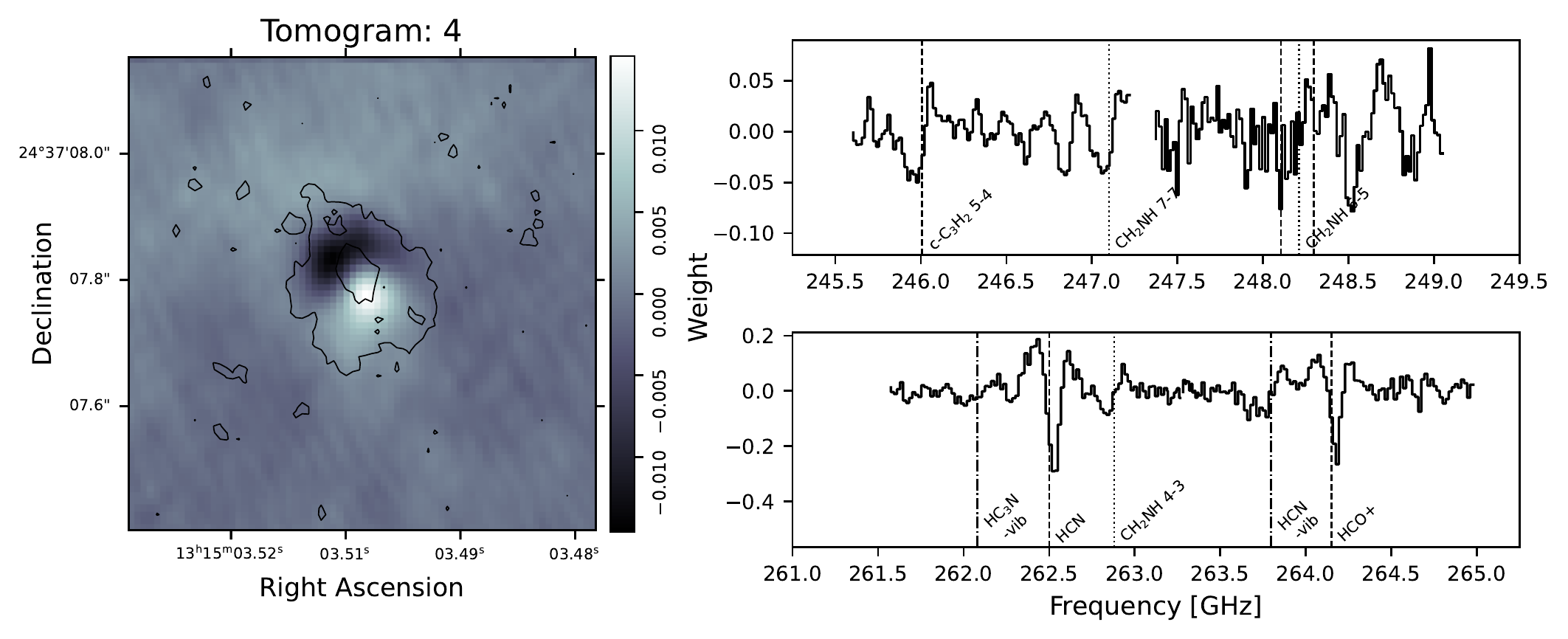}
    \includegraphics[width=0.99\textwidth]{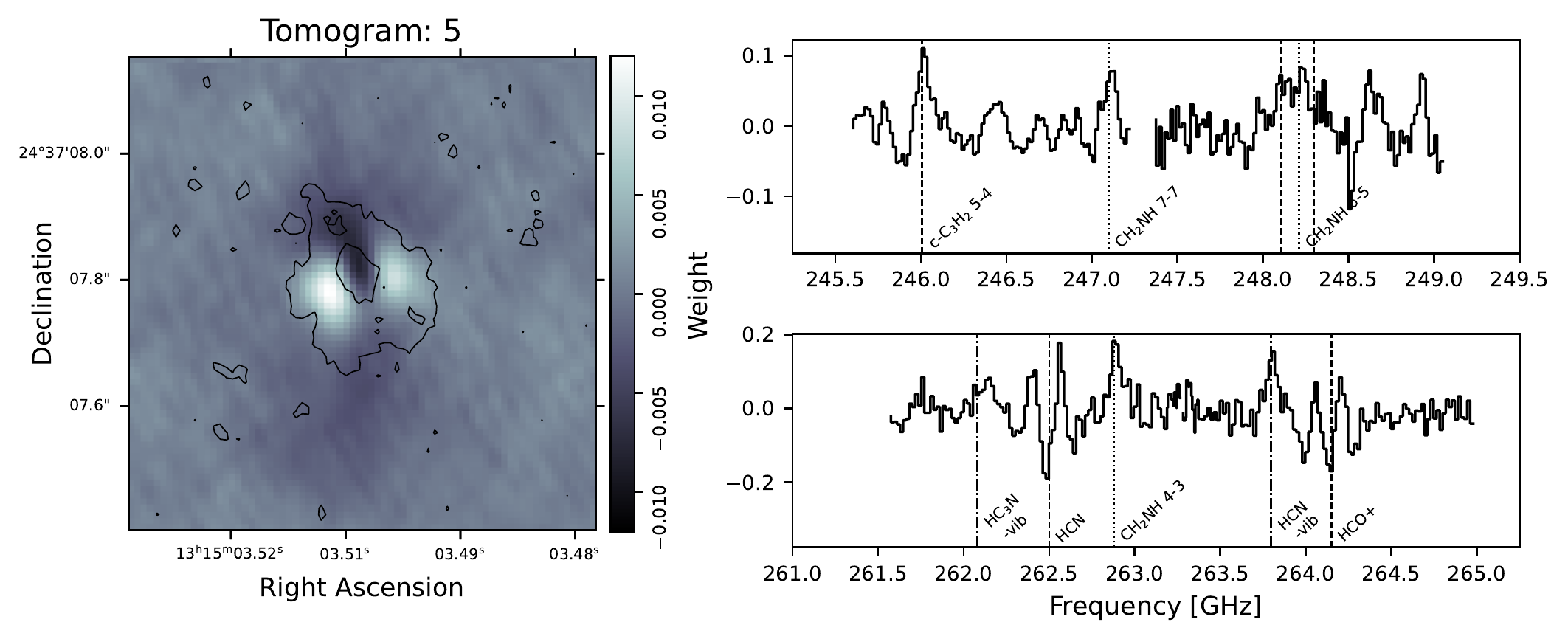}
    \includegraphics[width=0.99\textwidth]{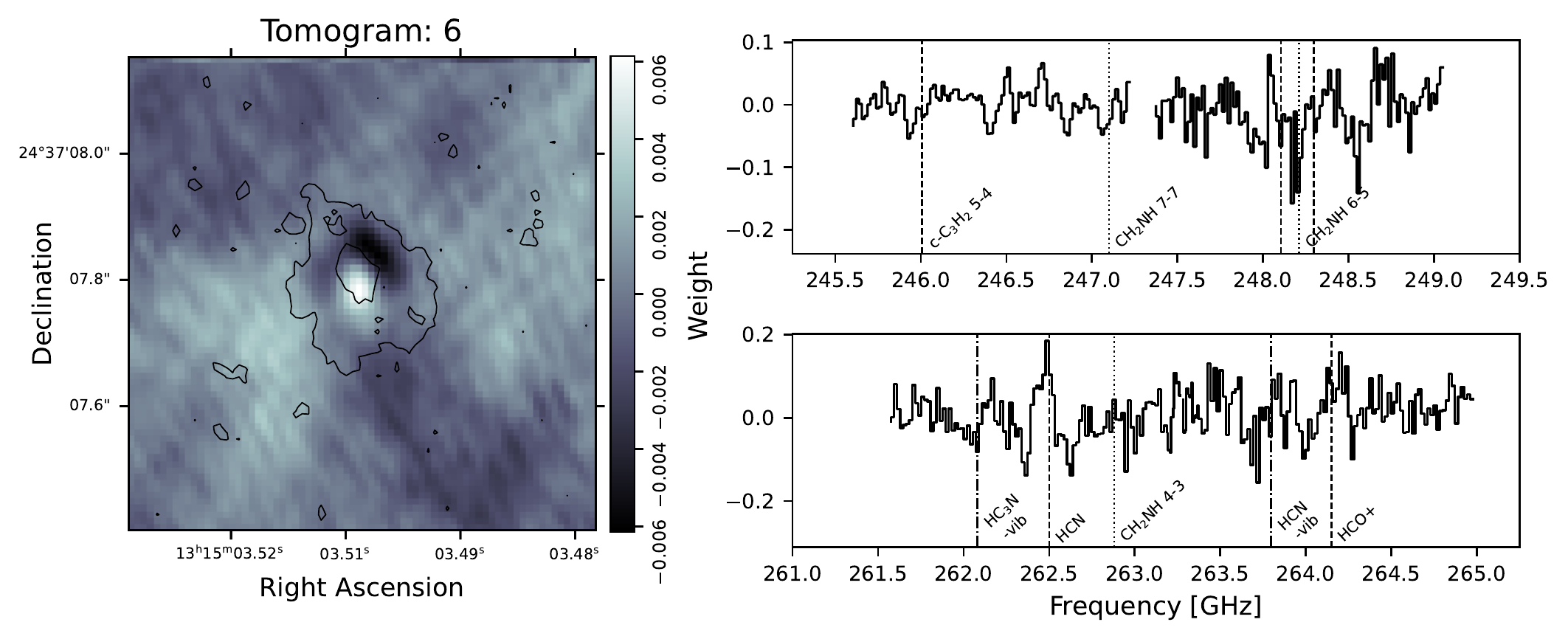}
    \caption{Principal components four to six. White means stronger correlation and black means anticorrelation. 
    The 3$\sigma$ contour of the CH$_2$NH $4_{13}$-$3_{12}$ transition is plotted in black.
    PC~4 shows anticorrelation between redshifted and blueshifted emission from vibrational transitions, indicating a more slowly rotating structure and a change in excitation of the molecular gas. 
    PC~5 shows a double-lobed structure associated with line wings.
    PC~6 shows a correlation with the low velocity components of the ground state HCN line and a weak anti-correlation with the line wings. 
    The vertical lines mark the redshifted rest frequencies of the same lines identified in Fig.~\ref{fig:PC0}.
    }
    \label{fig:PC4-6}
\end{figure*}

\begin{figure*}
    \centering
    \includegraphics[width=0.98\textwidth]{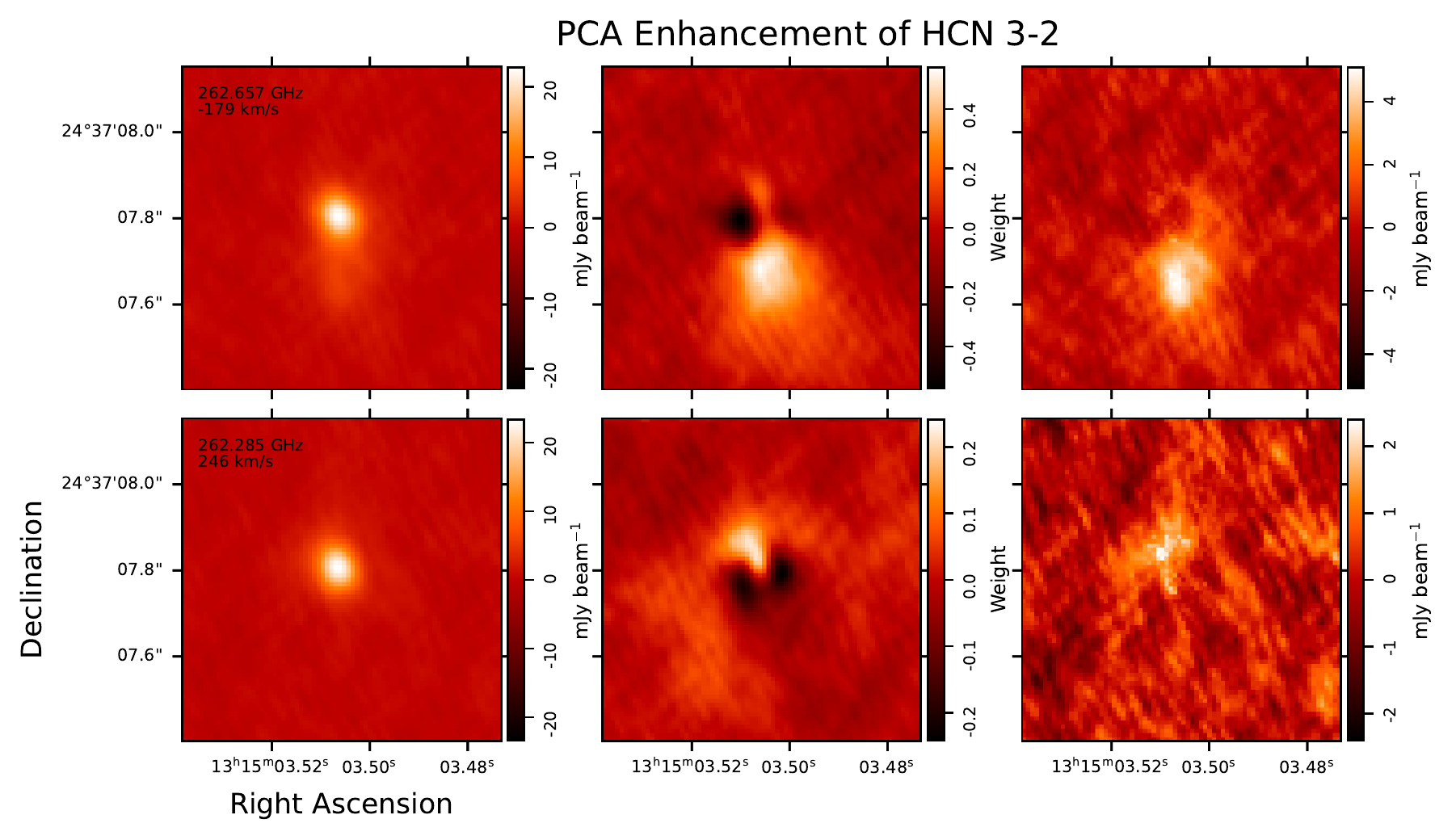}
    \caption{Comparison between the original data cube, the PCA-enhanced cube, and the continuum-subtracted data cube.
     The left column displays the original data cube, the center column the PCA-enhanced cube, and the right column the continuum-subtracted cube.
    The top row of panels shows the 262.657~GHz channel corresponding to -179~\kms  of the HCN~(4-3) transition, adopting a redshift of 0.0129 for IC~860. 
    The bottom row of panels shows the 262.285~GHz 246~\kms channel.
    The center image is given in weight and the other two panels in mJy~beam$^{-1}$.
    }
    \label{fig:PCAE-HCN}
\end{figure*}

\begin{figure*}
    \centering
    \includegraphics[width=0.98\textwidth]{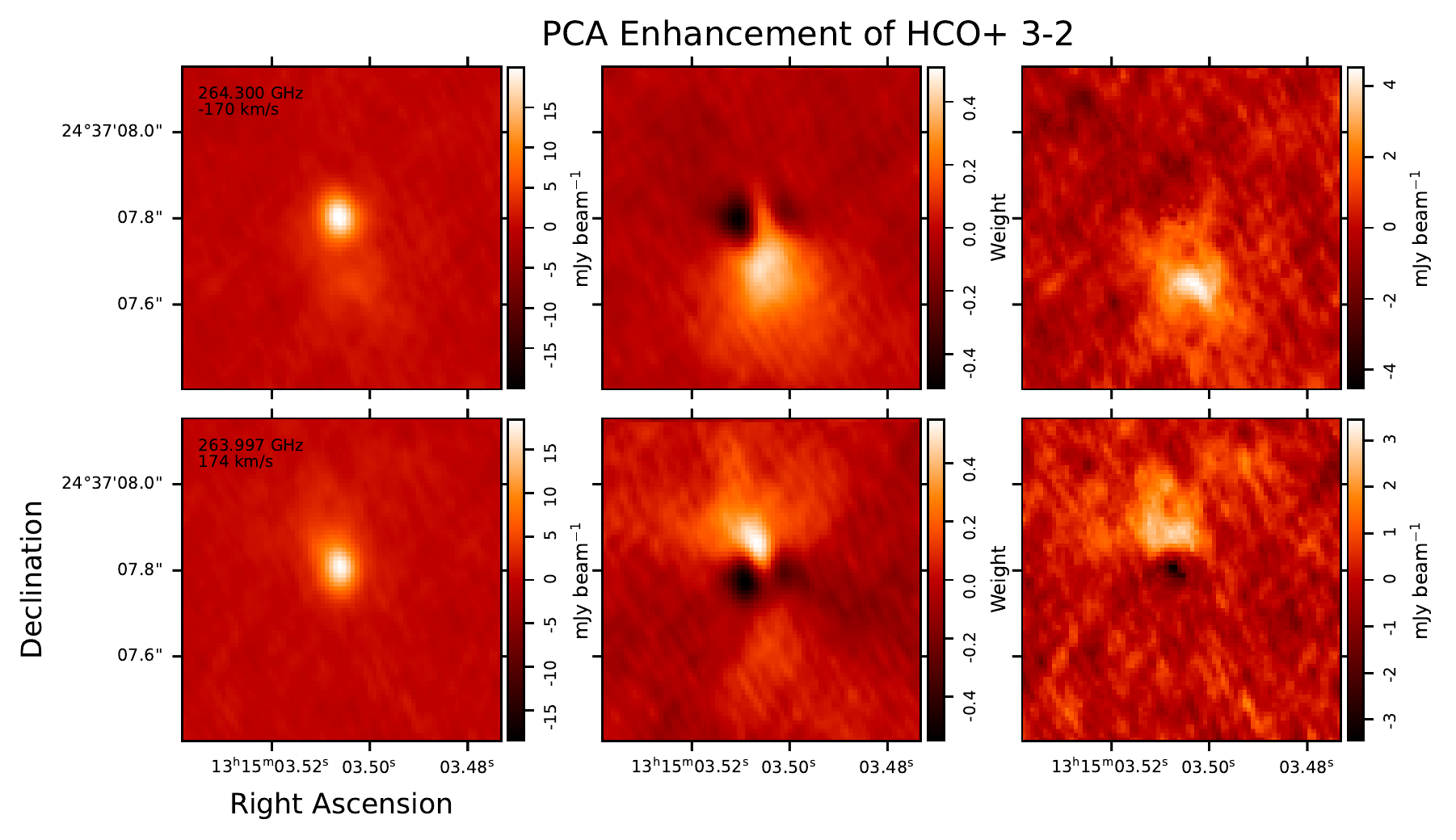}
    \caption{Comparison between the original data cube, the PCA-enhanced cube, and the continuum-subtracted data cube.
     The left column displays the original data cube, the center column the PCA-enhanced cube, and the right column the continuum-subtracted cube.
    The top row of panels shows the 264.318~GHz channel corresponding to -191~\kms  of the HCN~(4-3) transition, adopting a redshift of 0.0129 for IC~860. 
    The bottom row of panels shows the 263.979~GHz 194~\kms channel.
    The center image is given in weight and the other two panels in mJy~beam$^{-1}$.
    }
    \label{fig:PCAE-HCO+}
\end{figure*}

Figure \ref{fig:PC4-6} shows the tomograms and eigenvectors of IC~860. 
Figures \ref{fig:PCAE-HCN} and \ref{fig:PCAE-HCO+} show the comparison between the original data cubes, and the PCA-enhanced data cube. 
The left column shows channels from the original calibrated data cube. 
The center column shows a channel of the PCA-enhanced data cube. 
The right column shows channels of the continuum-subtracted data cube. 
The same channel, from the  three data cubes, is shown across each row and identified in the top right.

\end{appendix}

\end{document}